\documentclass[iop]{emulateapj}
\let\pwiflocal=\iffalse \let\pwifjournal=\iffalse

\usepackage{amsmath,amssymb}
\usepackage[breaklinks,colorlinks,urlcolor=blue,citecolor=blue,linkcolor=blue]{hyperref}

\usepackage[T1]{fontenc} 
\pwifjournal\else
  \usepackage{microtype}
\fi

\pwifjournal\else
  \makeatletter
  \renewcommand\plotone[1]{%
    \centering \leavevmode \setlength{\plot@width}{0.95\linewidth}
    \includegraphics[width={\eps@scaling\plot@width}]{#1}%
  }%
  \makeatother
\fi

\makeatletter

\newcommand\simpfx{http://simbad.u-strasbg.fr/simbad/sim-id?Ident=}

\newcommand\MakeObj[4][\@empty]{
  \pwifjournal%
    \expandafter\newcommand\csname pkgwobj@c@#2\endcsname[1]{\protect\object[#4]{##1}}%
  \else%
    \expandafter\newcommand\csname pkgwobj@c@#2\endcsname[1]{\href{\simpfx #3}{##1}}%
  \fi%
  \expandafter\newcommand\csname pkgwobj@f#2\endcsname{#4}%
  \ifx\@empty#1%
    \expandafter\newcommand\csname pkgwobj@s#2\endcsname{#4}%
  \else%
    \expandafter\newcommand\csname pkgwobj@s#2\endcsname{#1}%
  \fi}%

\newcommand\MakeTrunc[2]{
  \expandafter\newcommand\csname pkgwobj@t#1\endcsname{#2}}%

\newcommand{\obj}[1]{\csname pkgwobj@c@#1\endcsname{\csname pkgwobj@s#1\endcsname}}
\newcommand{\objf}[1]{\csname pkgwobj@c@#1\endcsname{\csname pkgwobj@f#1\endcsname}}
\newcommand{\objn}[1]{\csname pkgwobj@s#1\endcsname}
\newcommand{\objt}[1]{\csname pkgwobj@c@#1\endcsname{\csname pkgwobj@t#1\endcsname}}

\makeatother

\MakeObj[DENIS\,0255$-$4700]{dp0255}{DENIS-P\%20J025503.3-470049}{DENIS-P\,J025503.3$-$470049}
\MakeObj{g208-44}{G\%20208-44}{G\,208--44\,AB}
\MakeTrunc{g208-44}{G\,208--44}
\MakeObj{g208-45}{G\%20208-45}{G\,208--45}
\MakeObj{g208-44-45}{WDS\%20J19539\%2b4425AB}{G\,208-44AB/45}
\MakeObj{kelu1}{NAME\%20Kelu-1}{Kelu-1~AB}
\MakeObj{lhs248}{LHS\%20248}{LHS\,248}
\MakeObj{lhs292}{LHS\%20292}{LHS\,292}
\MakeObj{lhs523}{LHS\%20523}{LHS\,523}
\MakeObj{lhs2397}{LHS\%202397a}{LHS\,2397a\,AB}
\MakeTrunc{lhs2397}{LHS\,2397a}
\MakeObj{lhs2924}{LHS\%202924}{LHS\,2924}
\MakeObj{lhs3003}{LHS\%203003}{LHS\,3003}
\MakeObj{lhs3406}{LHS\%203406}{LHS\,3406}
\MakeObj{lp349}{LP\%20349-25}{LP\,349--25\,AB}
\MakeTrunc{lp349}{LP\,349--25}
\MakeObj{lp647}{LP\%20647-13}{LP\,647--13}
\MakeObj{lp851}{LP\%20851-346}{LP\,851--346}
\MakeObj{lp944}{LP\%20944-20}{LP\,944--20}
\MakeObj{luhman16}{2MASS\%20J10491891-5319100}{Luhman~16}
\MakeObj{n33370}{NLTT\%2033370}{NLTT\,33370\,AB}
\MakeTrunc{n33370}{NLTT\,33370}
\MakeObj{n40026}{NLTT\%2040026}{NLTT\,40026}
\MakeObj{tvlm513}{TVLM\%20513-46546}{TVLM\,513--46546}
\MakeObj{vb10}{vB\%2010}{vB\,10}
\MakeObj{0036+18}{2MASS\%20J00361617\%2b1821104}{2MASS\,J00361617$+$1821104}
\MakeObj{0746+20}{2MASS\%20J07464256\%2b2000321}{2MASS\,J07464256$+$2000321}
\MakeObj{1047+21}{2MASS\%20J10475385\%2b2124234}{2MASS\,J10475385$+$2124234}
\MakeObj{1048-39}{2MASSI\%20J1048147-395606}{2MASSI\,J1048147$-$395606}
\MakeObj{1835+32}{LSR\%20J1835\%2b3259}{LSR\,J1835$+$3259}

\newcommand\apx{\ensuremath{\sim}}
\newcommand\chandra{\textit{Chandra}}
\newcommand\cgsflux{erg~s$^{-1}$~cm$^{-2}$}

\newcommand\citeeg[1]{\citep[e.g.,][]{#1}}
\newcommand\ergps{erg~s$^{-1}$}
\newcommand\ha{{\ensuremath{\text{H}\alpha}}}
\newcommand\kms{km~s$^{-1}$}
\newcommand\msun{\ensuremath{M_\odot}}
\newcommand\pimms{\textsf{PIMMS}}

\newcommand\sherpa{\textsf{Sherpa}}
\newcommand\teff{\ensuremath{T_\text{eff}}}

\newcommand\ujy{$\mu$Jy}
\newcommand\ujybm{$\mu$Jy~bm$^{-1}$}

\newcommand\vsi{\ensuremath{v \sin i}}

\newcommand\Fx{\ensuremath{f_\text{X}}}
\newcommand\fx{\ensuremath{[\Fx]}}

\newcommand\Lb{\ensuremath{L_\text{bol}}}
\newcommand\Lh{\ensuremath{L_\ha}}
\newcommand\Lr{\ensuremath{L_\text{R}}}
\newcommand\sLr{\ensuremath{L_{\nu,\text{R}}}}
\newcommand\Lx{\ensuremath{L_\text{X}}}
\newcommand\Lxsat{\ensuremath{L_\text{X,sat}}}
\newcommand\lb{\ensuremath{[\Lb]}}

\newcommand\lr{\ensuremath{[\Lr]}}
\newcommand\slr{\ensuremath{[\sLr]}}
\newcommand\lx{\ensuremath{[\Lx]}}
\newcommand\LhLb{\ensuremath{\Lh/\Lb}}
\newcommand\LrLb{\ensuremath{\Lr/\Lb}}
\newcommand\sLrLb{\ensuremath{\sLr/\Lb}}
\newcommand\sLrLx{\ensuremath{\sLr/\Lx}}
\newcommand\LxLb{\ensuremath{\Lx/\Lb}}
\newcommand\lhlb{\ensuremath{[\LhLb]}}
\newcommand\lrlb{\ensuremath{[\LrLb]}}
\newcommand\slrlx{\ensuremath{[\sLrLx]}}
\newcommand\lxlb{\ensuremath{[\LxLb]}}

\defcitealias{paper2}{Paper~\textsc{ii}}
\newcommand\papertwo{\citetalias{paper2}}

\begin{document}

\title{Trends in Ultracool Dwarf Magnetism. I. X-Ray Suppression and Radio Enhancement}
\author{
  P.~K.~G. Williams\altaffilmark{1},
  B.~A. Cook\altaffilmark{2},
  and
  E. Berger\altaffilmark{1}
}
\email{pwilliams@cfa.harvard.edu}
\altaffiltext{1}{Harvard-Smithsonian Center for Astrophysics, 60 Garden Street,
  Cambridge, MA 02138, USA}
\altaffiltext{2}{Department of Astrophysical Sciences, Princeton University,
  Princeton, NJ 08544, USA}

\slugcomment{Draft: \today}
\shorttitle{UCD Magnetism Trends. I. X-ray Suppression, Radio Enhancement}
\shortauthors{Williams \textit{et al.}}

\begin{abstract}
  Although ultracool dwarfs (UCDs) are now known to generate and dissipate
  strong magnetic fields, a clear understanding of the underlying dynamo is
  still lacking. We have performed X-ray and radio observations of seven UCDs
  in a narrow range of spectral type (M6.5--M9.5) but spanning a wide range of
  projected rotational velocities ($\vsi \approx 3$--40~\kms). We have also
  analyzed unpublished archival \chandra\ observations of four additional
  objects. All of the newly-observed targets are detected in the X-ray, while
  only one is detected in the radio, with the remainder having sensitive upper
  limits. We present a database of UCDs with both radio and X-ray measurements
  and consider the data in light of the so-called G\"udel-Benz relation (GBR)
  between magnetic activity in these bands. Some UCDs have very bright radio
  emission and faint X-ray emission compared to what would be expected for
  rapid rotators, while others show opposite behavior. We show that UCDs would
  still be radio-over-luminous relative to the GBR even if their X-ray
  emission were at standard rapid-rotator ``saturation'' levels. Recent
  results from Zeeman-Doppler imaging and geodynamo simulations suggest that
  rapidly-rotating UCDs may harbor a bistable dynamo that supports either a
  stronger, axisymmetric magnetic field or a weaker, non-axisymmetric field.
  We suggest that the data can be explained in a scenario in which
  strong-field objects obey the GBR while weak-field objects are
  radio-over-luminous and X-ray-under-luminous, possibly because of a
  population of gyrosynchrotron-emitting coronal electrons that is
  continuously replenished by low-energy reconnection events.
\end{abstract}


\section{Introduction}
\label{s.intro}

The magnetic activity of very low mass stars and brown dwarfs (collectively,
``ultracool dwarfs'' or UCDs; here, objects of spectral types \apx M7 or
later) raises several challenging astrophysical questions. UCD activity cannot
be driven by a solar-type shell dynamo, in which the tachocline (the shearing
interface layer between the radiative and convective zones) plays a crucial
role; full convection sets in around spectral type M4, and so the tachocline
simply does not exist in UCDs \citep{cb00}. Nonetheless, observations reveal
persistent, strong (\apx kG) magnetic fields at the bottom of the main
sequence \citep{bbb+01, b02, b06b, bbf+10, had+06, rb07, opbh+09, rb10,
  mdp+10, ahd+13}. Despite substantial theoretical work to understand the
fully-convective dynamo \citeeg{ddyr93,ck06,dsb06,b08}, its properties, and
those of the resulting magnetic fields, remain unclear. Further progress
requires observational input based on studies of magnetic activity indicators.

X-ray and \ha\ emission are two common indicators, and decades of study of
solar-type stars have yielded several important relationships between these
quantities and other stellar properties. One such relationship is between
X-ray activity (\LxLb, where the denominator is the bolometric stellar
luminosity) and stellar rotation \citep {nhb+84}, which follows a
``saturation'' pattern in which activity increases with rotation until it
reaches $\LxLb \sim 10^{-3}$ \citep{v84,pmm+03,jjb+11,wdmh11}. The same
general pattern is observed in \ha\ emission \citeeg{dfpm98}.

Radio emission is another powerful magnetic activity indicator, and there are
also well-known correlations between the radio and soft X-ray (SXR)
luminosities of magnetically active stellar systems. \citet{dsl89} first noted
a correlation betwen \sLr\ and \Lx\ in RS\,CVn systems. \citet{gb93} analyzed
a larger sample and found that for the most active F--M stars, $\Lx \sim
\sLr\times10^{15.5}$~Hz. This result was then extended to solar flares and
other active binaries by \citet{bg94}. Over their whole dataset, spanning 10
orders of magnitude in radio spectral luminosity, $\Lx \propto \sLr^\alpha$
with $\alpha \sim 0.73$, a result now commonly known as the G\"udel-Benz
relation (GBR). Continuity over such a broad range strongly suggests both a
common driver of emission in the two bands (despite the fact that the
fundamental emission processes operate in very different conditions), as well
as common physical processes across this range of emitters. In the standard
interpretation, magnetic reconnection accelerates a population of nonthermal
particles, leading to radio emission; these particles then deposit some of
their energy in the chromosphere, where ablated material concentrates in
coronal loops and emits thermally in the SXR band. This model is supported by
the observed ``Neupert effect'' \citep{n68}, in which $\text{d}\Lx/\text{d}t
\propto \Lr$, suggesting that the SXR emission tracks the total energy
deposited by the particle acceleration process. This effect is
well-established though far from universal in both the solar and stellar
contexts \citeeg{dz93,gbss96,oba+04}.

Observations of magnetic activity tracers in UCDs paint a different picture.
Despite the demonstrated existence of strong fields in at least some UCDs,
their X-ray and \ha\ emission both drop off precipitously \citep{smf+06,
  bbf+10, gmr+00, whw+04}. The rotation/activity relation evolves
significantly in UCDs, with evidence for trends in which X-ray and
\ha\ activity decrease as rotation increases \citep{bm95, mb03, wb09, bbf+10},
reminiscent of the much weaker ``supersaturation'' effect seen in active stars
\citep{prs+96,jjb+11}. But UCD radio emission remains stubbornly unchanged: in
this regime, radio activity and radio surface flux ($L_\text{rad}/R_*^2$)
increase with rotation, with no evidence of saturation \citep{mbr12}.

These trends offer clues toward a deeper understanding of the fully convective
dynamo and the structures of (sub)stellar magnetic fields and outer
atmospheres. Progress toward this understanding, however, is hampered by the
relatively small number of UCDs detected in the X-ray regime. In an attempt to
improve this situation, we have observed numerous UCDs with the
\chandra\ X-ray Observatory \citep{brr+05, bgg+08, bbg+08, brpb+09, bbf+10}.
In this work, we report new \chandra\ observations of 7 UCDs of spectral type
\apx M7 with a wide range of rotational velocities, all of which were
detected. Every source was also observed (non-simultaneously) with the
upgraded Karl G. Jansky Very Large Array (VLA), yielding one detection. Our
\chandra\ detections offer a striking counterpoint to previous observations of
objects objects only a few spectral subtypes later, in which X-ray detections
have been elusive \citeeg{brr+05,bbf+10}. We also report the analysis of
several unpublished measurements from archival data. We combine these results
with data from our previous work and the literature to form a comprehensive
database of UCD activity measurements.

In this work, we use this database to investigate the correlation between
X-ray and radio emission in UCDs. We proceed here by describing the targets
for which we present new results (\S\ref{s.targets}) and our data analysis
(\S\S\ref{s.xray}, \ref{s.radio}). We then discuss our results in the context
of the full sample of UCDs with X-ray and radio observations
(\S\ref{s.trends}). We offer a physical model that explains the observations
(\S\ref{s.disc}). In \citet[hereafter \papertwo]{paper2}, we use the same
database to investigate the correlation between UCD X-ray emission and
rotation.

Throughout this work, we use the notation $[x] \equiv \log_{10} x$, with $x$
being measured in cgs units if it is a dimensional quantity, unless its units
are specified otherwise.

\section{Targets with New Results}
\label{s.targets}

Candidates for new radio and X-ray observations were selected by searching
\href{http://dwarfarchives.org}{dwarfarchives.org} for nearby UCDs of spectral
type \apx M7 that were visible to the VLA and had measurements of \vsi\ in the
literature. We also searched the \chandra\ data archive for unpublished
observations of late-type objects with \vsi\ measurements. Seven targets were
observed with \chandra\ and the VLA, and we identified and analyzed four
archival targets. The main characteristics of these objects, along with the
appropriate references and 2MASS identifications \citep{the2mass}, are
provided in Table~\ref{t.tinfo}, while their properties are discussed in
greater detail in the Appendix. The Appendix also describes our method for
computing bolometric luminosities.

\begin{deluxetable*}{cllr@{}l@{\,}lcr@{}l@{\,}lc@{ }c@{ }c@{ }c@{ }c@{ }c}

\tablecolumns{16}
\tablewidth{0em}
\tablecaption{Properties of Ultracool Dwarfs with New Analysis Presented in This Paper\label{t.tinfo}}
\tablehead{
\colhead{~~} & \colhead{2MASS Identifier} & \colhead{Name} & \multicolumn{3}{c}{Distance} & \colhead{SpT} & \multicolumn{3}{c}{\vsi} & \multicolumn{6}{c}{References}      \\
 \cline{11-16}  &  &  & \multicolumn{3}{c}{(pc)} &  & \multicolumn{3}{c}{(\kms)} &  &  &  &  &  &  \\ \\
\multicolumn{1}{c}{} & \multicolumn{1}{c}{(1)} & \multicolumn{1}{c}{(2)} & \multicolumn{3}{c}{(3)} & \multicolumn{1}{c}{(4)} & \multicolumn{3}{c}{(5)} & \multicolumn{1}{c}{(P)} & \multicolumn{1}{c}{(C)} & \multicolumn{1}{c}{(M)} & \multicolumn{1}{c}{(D)} & \multicolumn{1}{c}{(S)} & \multicolumn{1}{c}{(V)}
}
\startdata
\sidehead{\textit{Newly-observed targets}} & 01095117$-$0343264 & \obj{lp647}\tablenotemark{a} & $11$ & $.1$ & $\pm\,0.7$\rule{0pt}{3ex} & M9 & $13$ &  & $\pm\,2$\rule{0pt}{3ex} & 1 & 2 &  & 3 & 3 & 4 \\
 & 10481258$-$1120082 & \obj{lhs292} & $4$ & $.54$ & $\pm\,^{0.06}_{0.07}$\rule{0pt}{3ex} & M6.5 & \multicolumn{3}{c}{$<$$3$}\rule{0pt}{3ex} & 5 & 6 &  & 8 & 7 & 4 \\
 & 11214924$-$1313084\,A & \objt{lhs2397}\,A & $14$ & $.3$ & $\pm\,^{0.5}_{0.4}$\rule{0pt}{3ex} & M8 & $15$ &  & $\pm\,1$\rule{0pt}{3ex} & 5 & 9 & 10 & 8 & 10 & 11 \\
 & \phantom{11214924$-$1313084\,}B & \phantom{\objt{lhs2397}}\,B &  & & \rule{0pt}{3ex} & L7.5 & $11$ &  & $\pm\,3$\rule{0pt}{3ex} &  &  &  &  & 10 & 11 \\
 & 11554286$-$2224586 & \obj{lp851} & $9$ & $.7$ & $\pm\,1.0$\rule{0pt}{3ex} & M7.5 & $33$ &  & $\pm\,3$\rule{0pt}{3ex} & 5 & 12 &  & 13 & 13 & 4 \\
 & 15210103$+$5053230 & \obj{n40026} & $16$ &  & $\pm\,2$\rule{0pt}{3ex} & M7.5 & $40$ &  & $\pm\,4$\rule{0pt}{3ex} & 1 & 3 &  & 3 & 3 & 4 \\
 & 18432213$+$4040209 & \obj{lhs3406} & $14$ & $.1$ & $\pm\,0.2$\rule{0pt}{3ex} & M8 & $5$ &  & $\pm\,3.2$\rule{0pt}{3ex} & 5 & 14 &  & 8 & 3 & 4 \\
 & 22285440$-$1325178 & \obj{lhs523} & $11$ & $.3$ & $\pm\,0.6$\rule{0pt}{3ex} & M6.5 & $7$ & $.0$  & $\pm\,2$\rule{0pt}{3ex} & 5 & 15 &  & 8 & 16 & 17 \\
\sidehead{\textit{\chandra\ archival targets}} & 00275592$+$2219328\,A & \objt{lp349}\,A & $13$ & $.2$ & $\pm\,0.3$\rule{0pt}{3ex} & M8 & $55$ &  & $\pm\,2$\rule{0pt}{3ex} & 5 & 18 & 19 & 20 & 19 & 11 \\
 & \phantom{00275592$+$2219328\,}B & \phantom{\objt{lp349}}\,B &  & & \rule{0pt}{3ex} & M9 & $83$ &  & $\pm\,3$\rule{0pt}{3ex} &  &  &  &  & 19 & 11 \\
 & 02550357$-$4700509 & \obj{dp0255} & $4$ & $.98$ & $\pm\,0.09$\rule{0pt}{3ex} & L8 & $67$ &  & $\pm\,13$\rule{0pt}{3ex} & 21 & 22 &  & 23 & 3 & 24 \\
 & 19535443$+$4424541\,A & \objt{g208-44}\,A & $4$ & $.61$ & $\pm\,0.04$\rule{0pt}{3ex} & M5.0 & $22$ & $.5$  & $\pm\,2$\rule{0pt}{3ex} & 5 & 25 & 26 & 28 & 27 & 17 \\
 & \phantom{19535443$+$4424541\,}B & \phantom{\objt{g208-44}}\,B &  & & \rule{0pt}{3ex} & M8.5 & $17$ & $.4$  & $\pm\,1.4$\rule{0pt}{3ex}\tablenotemark{b} & 5 & 29 &  &  & 27 & 28 \\
 & 19535508$+$4424550 & \obj{g208-45} &  & & \rule{0pt}{3ex} & M5.5 & $6$ & $.8$  & $\pm\,1.9$\rule{0pt}{3ex} & 5 & 25 &  &  & 14 & 28
\enddata
\tablecomments{See Appendix for additional details and references.}
\tablerefs{Columns are (P), discovery of substantial proper motion;
(C), classification as very cool dwarf; (M), discovery of multiplicity; (S),
spectral type; (D), distance; (V), \vsi. [1] \citet{thenltt}, [2] \citet{cr02}, [3] \citet{crl+03}, [4] \citet{rb10}, [5] \citet{thelhs}, [6] \citet{dlh86}, [7] \citet{hks94}, [8] \citet{thegctp}, [9] \citet{thethirdgj}, [10] \citet{fcs03}, [11] \citet{kgf+12}, [12] \citet{pbcd+03}, [13] \citet{cpbd+05}, [14] \citet{rhg95}, [15] \citet{ldgs79}, [16] \citet{khm91}, [17] \citet{mb03}, [18] \citet{gmr+00}, [19] \citet{fbd+05}, [20] \citet{gc09}, [21] \citet{cjb08}, [22] \citet{mdb+99}, [23] \citet{cmj+06}, [24] \citet{rb08}, [25] \citet{hdg74}, [26] \citet{hd84}, [27] \citet{lhm08}, [28] \citet{dfpm98}, [29] \citet{mhf+88}}
\tablenotetext{a}{Possible binary; see \citet{gw03}.}
\tablenotetext{b}{Measurement is a blend of both system components.}
\end{deluxetable*}

\subsection{Observations}
\label{s.obs}

The new observations were performed with \chandra/ACIS-S between 2011~December
and 2013~February (proposal 13200167; \chandra\ observation IDs 13603--13609;
PI: Berger), using the S3 backside-illuminated chip. All exposures were 20~ks,
except for \obj{lhs292} which was observed for 10~ks. Parameters of the
observations are provided in Table~\ref{t.xinfo}. No grating was used, the
data mode was \textsf{VFAINT}, and the exposure mode was ``timed'' (TE).

\begin{deluxetable*}{clccccc}

\tablecolumns{7}
\tablewidth{0em}
\tablecaption{Parameters of \chandra\ Observations\label{t.xinfo}}
\tablehead{
\colhead{~~} & \colhead{Name} & \multicolumn{2}{c}{Observation Date}  & \colhead{ObsId} & \colhead{Integ. Time} & \colhead{Raw Counts} \\
 \cline{3-4}  &  & \colhead{Gregorian} & \colhead{MJD[TT]} &  & \colhead{(s)} &  \\ \\
\multicolumn{1}{c}{} & \multicolumn{1}{c}{(1)} & \multicolumn{1}{c}{(2)} & \multicolumn{1}{c}{(3)} & \multicolumn{1}{c}{(4)} & \multicolumn{1}{c}{(5)} & \multicolumn{1}{c}{(6)}
}
\startdata
\sidehead{\textit{Newly-observed targets}} & \obj{lhs292} & 2011 Dec 14 & $55909.56$ & 13603 & $10620$ & 51 \\
 & \obj{lhs523} & 2012 Dec 08 & $56269.18$ & 13604 & $19605$ & 15 \\
 & \obj{lhs2397} & 2012 Nov 05 & $56236.01$ & 13606 & $19804$ & 252 \\
 & \obj{lhs3406} & 2012 Dec 05 & $56266.17$ & 13609 & $19804$ & 74 \\
 & \obj{lp647} & 2012 Oct 24 & $56224.21$ & 13605 & $19798$ & 12 \\
 & \obj{lp851} & 2012 Jul 09 & $56117.63$ & 13607 & $19800$ & 43 \\
 & \obj{n40026} & 2013 Feb 07 & $56330.68$ & 13608 & $18812$ & 6 \\
\sidehead{\textit{\chandra\ archival targets}} & \obj{g208-44} & 2003 Dec 14 & $52987.18$ & 4476 & $23785$\tablenotemark{a} & 1607\tablenotemark{b} \\
 & \obj{g208-45} & 2003 Dec 14 & $52987.18$ & 4476 & $23785$\tablenotemark{a} & 1072\tablenotemark{b} \\
 & \obj{dp0255} & 2008 Dec 02 & $54802.28$ & 8903 & $17993$ & 0 \\
 & \obj{dp0255} & 2008 Dec 04 & $54804.79$ & 10828 & $9914$ & 1 \\
 & \obj{lp349} & 2009 Sep 15 & $55089.28$ & 9225 & $37197$ & 60
\enddata
\tablecomments{Cols. (2) and (3) are the approximate midpoint of the on-source
integration. Col. (4) is the \chandra\ observation ID. Col. (5) is the
on-source integration time, not accounting for the background flares described
in the text. Col. (6) is the number of counts in a 2$''$ aperture around the
predicted source position after processing in the \textsf{VFAINT} background
mode, not accounting for pileup.}
\tablenotetext{a}{Affected by background flaring (see \S\ref{s.xray}).}
\tablenotetext{b}{Affected by pileup (see \S\ref{s.xray}).}
\end{deluxetable*}

The targets were also observed with the Karl G. Jansky Very Large Array (VLA)
between 2012~February and 2012~May (project VLA/12A-089; PI: Berger). These
observations were not simultaneous with the X-ray observations, and the
relative timing between observations in the two bands was arbitrary. Each
observing session lasted 1~hr, with a total correlated bandwidth of 2048 MHz
divided into two basebands centered at 5000 and 7100 MHz, each containing 512
spectral channels. Bandpass, flux density scale, and complex gain calibrations
were obtained in the usual way; the sources used are listed in
Table~\ref{t.rinfo}, along with other parameters of the observations.
Approximately 65\% of each session was spent integrating on the target source.

\begin{deluxetable*}{lcccccc}

\tablecolumns{7}
\tablewidth{0em}
\tablecaption{Parameters of VLA Observations\label{t.rinfo}}
\tablehead{
\colhead{Name} & \multicolumn{2}{c}{Observation Date}  & \colhead{Integ. Time} & \colhead{Config.} & \colhead{Flux Cal.} & \colhead{Gain Cal.} \\
 \cline{2-3}  & \colhead{Gregorian} & \colhead{MJD[TT]} & \colhead{(s)} &  &  &  \\ \\
\multicolumn{1}{c}{(1)} & \multicolumn{1}{c}{(2)} & \multicolumn{1}{c}{(3)} & \multicolumn{1}{c}{(4)} & \multicolumn{1}{c}{(5)} & \multicolumn{1}{c}{(6)} & \multicolumn{1}{c}{(7)}
}
\startdata
\obj{lhs292} & 2012 Feb 22 & $55979.24$ & $2260$ & C & 3C\,286 & J1039$-$1541 \\
\obj{lhs523} & 2012 Mar 17 & $56003.76$ & $2205$ & C & 3C\,48 & J2246$-$1206 \\
\obj{lhs523} & 2012 May 21 & $56068.50$ & $2200$ & CnB & 3C\,48 & J2246$-$1206 \\
\obj{lhs2397} & 2012 Feb 22 & $55979.36$ & $2335$ & C & 3C\,286 & J1130$-$1449 \\
\obj{lhs3406} & 2012 Mar 17 & $56003.71$ & $2205$ & C & 3C\,48 & J1845$+$4007 \\
\obj{lp647} & 2012 Feb 22 & $55979.04$ & $2265$ & C & 3C\,48 & J0110$-$0741 \\
\obj{lp851} & 2012 Feb 28 & $55985.27$ & $2205$ & C & 3C\,286 & J1159$-$2148 \\
\obj{n40026} & 2012 Mar 16 & $56002.54$ & $2205$ & C & 3C\,286 & J1545$+$5135
\enddata
\tablecomments{\objn{lhs523} was observed twice. Cols. (2) and (3) are the approximate
midpoint of the on-source integration. Col. (4) is the on-source integration
time. Col. (5) is the configuration of the VLA at the time of the
observation.}
\end{deluxetable*}

The archival data were obtained as follows. \obj{g208-44-45} was observed with
\chandra/ACIS-S on 2003~December~14 (proposal 05200058; \chandra\ observation
ID 4476; PI: Garmire) for 24~ks. \obj{dp0255} was observed in two parts
(proposal 09200200; observation IDs 8903 and 10828; PI: Audard) on
2008~December~2 and 2008~December~4, for a total exposure time of 28~ks.
\obj{lp349} was observed on 2009~September~15 (proposal 10200468; observation
ID 9925; PI: Osten) with an exposure time of 37~ks. In all cases, the
instrumental configuration was the same as in our new observations.

\section{X-Ray Analysis}
\label{s.xray}

We analyzed the \chandra\ data in CIAO version 4.5 \citep{theciao} with CalDB
version 4.5.5.1. We used the date of each observation and astrometric
information from the Simbad database to predict positions for each target at
the time of our observations. Based on Monte Carlo simulations, all of our
predictions have uncertainties of $\lesssim2''$, and most of them have
uncertainties of $\lesssim0.5''$. This is comparable to the astrometric
precision of \chandra. We defined initial source apertures $2''$ in radius
centered on our astrometric predictions of the source positions. As discussed
below, two of our targets (\obj{g208-44} and \obj{g208-45}) are affected by
pileup, leading us to use annular apertures instead. For the rest of the
targets, the initial apertures needed no modification.

Following \textsf{VFAINT} reprocessing to eliminate a substantial fraction of
the background events, we estimated the mean residual background in each
dataset by extracting events in an energy range of 0.3--7~keV in large,
source-free regions near the target locations. In all cases, the expected
number of background counts in the source aperture is $\lesssim$1.

\begin{figure}[htbp]
\plotone{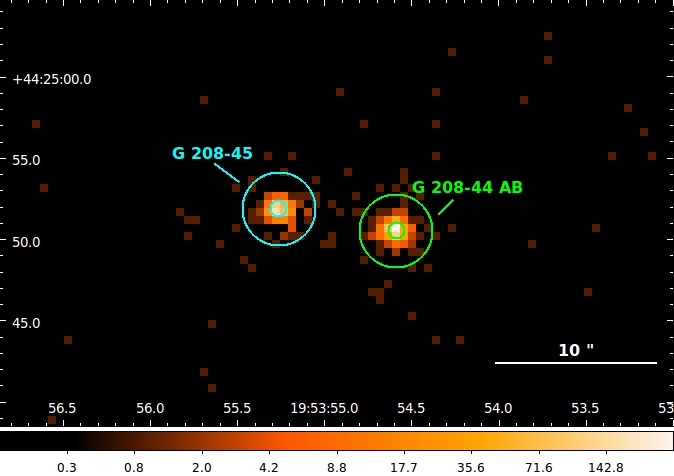}
\caption{Annotated \chandra\,image cutout of the \objn{g208-44-45} system.
  This image was generated after filtering out periods of background flaring
  as described in \S\ref{s.g208}. \objn{g208-45} is to the east (left) and the
  unresolved pair \objn{g208-44} is to the west (right). The annuli around
  each object represent the areas from which counts were extracted to avoid
  pileup in the central pixels (\S\ref{s.g208}). A colorbar below indicates
  the scale, which is in terms of total counts accumulated over the unfiltered
  intervals of the observation.}
\label{f.g208img}
\end{figure}

\begin{figure*}[htbp]
  \plotone{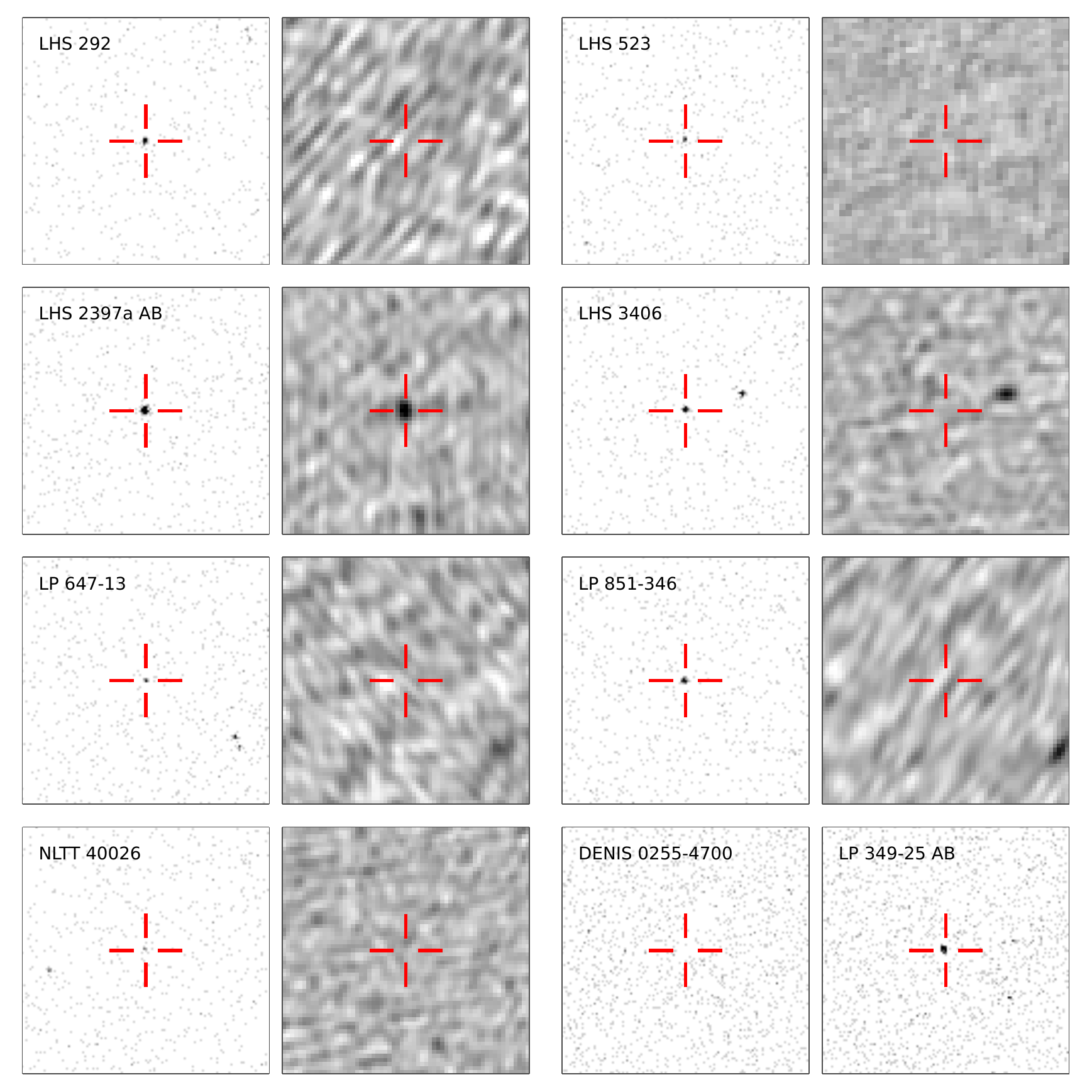}
  \caption{X-ray and radio image cutouts of all sources except
    \obj{g208-44-45} (Figure~\ref{f.g208img}). Each panel is $1'$$\times$$1'$.
    In each pair of panels except the last, the left-hand panel shows X-ray
    counts after \textsf{VFAINT} background processing on a white-to-black
    linear scale ranging from 0 to 10 counts. The right-hand panels show radio
    brightness on a white-to-black linear scale ranging from -20 to 50 \ujybm.
    Radio results from the literature were used for \obj{dp0255} and
    \obj{lp349}, which are shown as the bottom-right pair of panels. The
    radio/X-ray source near the predicted position of \obj{lhs3406} is
    discussed in \S\ref{s.radio}. Note also the plausible joint radio/X-ray
    detections of unrelated sources within the cutouts of \obj{lp647} (36
    X-ray counts, combining the resolved pair; $S_{\nu,\text{max}} \approx
    28$~\ujybm) and \obj{lp851} (4 counts within a $1''$ aperture;
    $S_{\nu,\text{max}} \approx 43$~\ujybm).}
  \label{f.cutouts}
\end{figure*}

Every source except the archival L8~dwarf \obj{dp0255} was detected at
$>$5$\sigma$ significance. In our new observations, the number of counts at
the source location ranges between 6 (\obj{n40026}, with $0.17$ expected
background events) and 252 (\obj{lhs2397}). With the high success rate in our
new observations, we nearly double the number of UCDs with X-ray detections
(cf. Table~\ref{t.dbsrcs}). The only target with nontrivial X-ray structure
and annular source extraction apertures, the triple system \obj{g208-44-45},
is shown in Figure~\ref{f.g208img}. Cutouts of the X-ray images around the
predicted locations of the other sources (as well as radio images; see next
section) are shown in Figure~\ref{f.cutouts}.

During extraction we also checked for background flaring. The only dataset in
which flares were seen was the observation of \obj{g208-44-45}. We discuss our
handling of this dataset separately below.

We searched for variability in the X-ray emission of the detected sources
using a Bayesian blocks analysis \citep{s98,snjc13}. This approach models the
source flux as a series of independent, piecewise constant ``blocks,'' with
overfitting being controlled by the use of a downward-sloping prior on the
number of blocks ($N_b$). Our implementation of the algorithm uses the
iterative approach described in \citet{snjc13} with a Monte-Carlo-derived
parametrization of the prior on $N_b$ that sets the probability of false
detection of an extraneous block at 5\%. This parametrization is given in
Equation~21 of \citet{snjc13} but is misstated; the correct equation is
\begin{equation}
\text{ncp\_prior} = 4 - \ln \left(73.53 p_0 N^{-0.478}\right),
\end{equation}
as may be verified with the example given in that work. Our implementation is
in Python and derives from the MatLab\texttrademark\ code provided by
\citet{snjc13} and the \textsf{AstroML} Python module \citep{theastroml}.
Compared to the latter module, our system adds support for time-tagged events
and datasets with gaps in coverage (due to the background flaring in our
case). It also fixes several minor bugs such as the mistaken equation above.
Our implementation is publicly available\footnote{The implementation is
  versioned using Git and is currently available at
  \url{https://github.com/pkgw/pwpy/blob/master/scilib/xbblocks.py}. The
  version used in this work is that included in commit
  \textsf{606218c4\-7d667e97\-d58c38f1\-fd09e7dc5\-540b38f}. The design of Git
  ensures that this commit identifier uniquely specifies the exact content and
  complete revision history of the code in question.}.

\begin{figure*}[hbt]
  \plotone{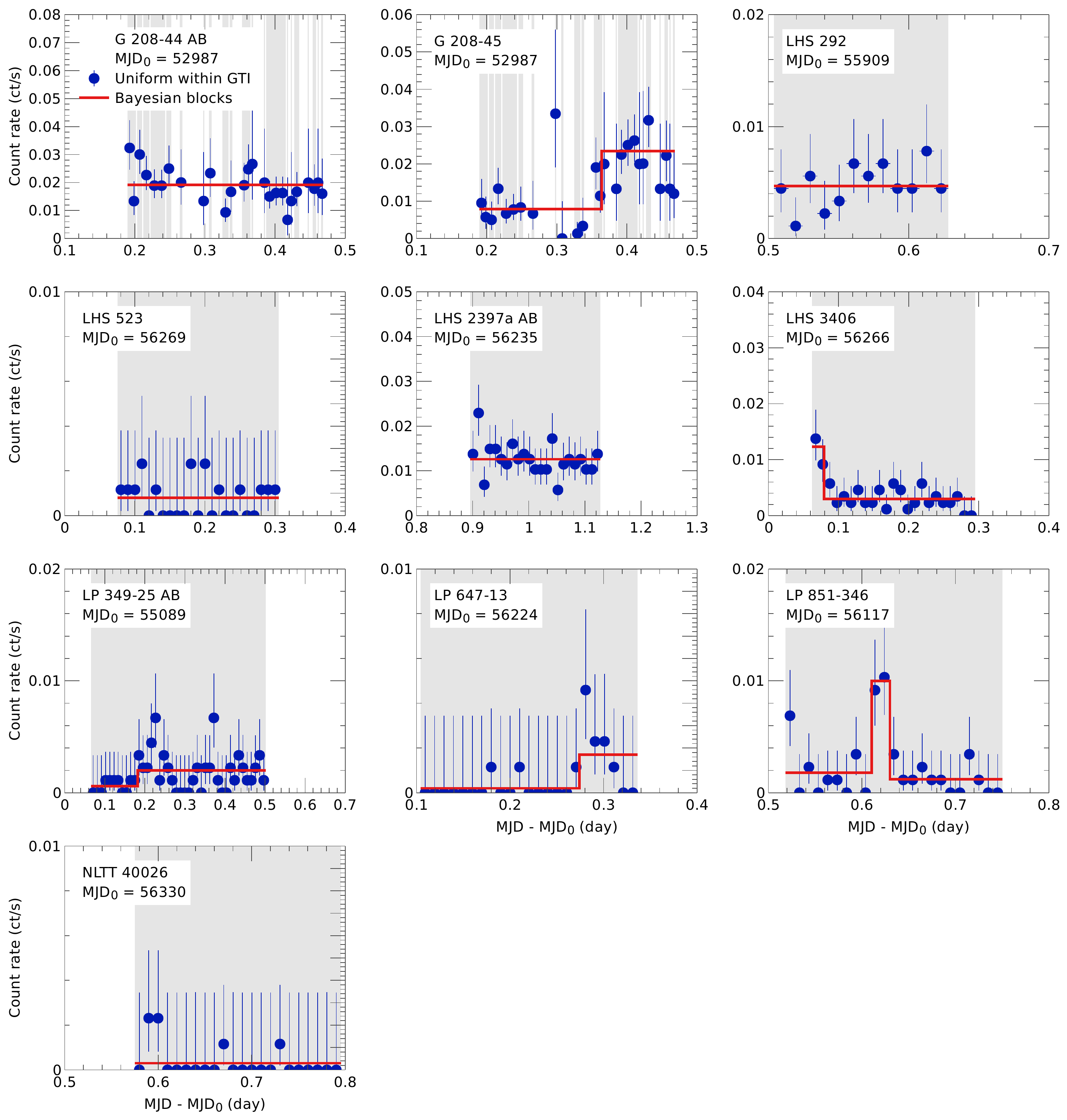}
  \caption{Light curves of the detected \chandra\ targets in the 0.3--7~keV
    energy range. Note the differing axis ranges for each panel.
    \textit{Shaded regions} indicate ``good times'' in which background flares
    are not dominant (\S\ref{s.g208}). \textit{Blue points} give count rates
    derived from bins that are uniformly-sized within each good time interval,
    with error bars from Poisson statistics. \textit{Red histograms} give the
    piecewise constant model derived from Bayesian block variability analysis
    (\S\ref{s.xray}).}
  \label{f.gtilcs}
\end{figure*}

The Bayesian blocks analysis finds more than one block --- that is,
significant evidence of variability --- in five sources. We plot the X-ray
light curves resulting from this analysis in Figure~\ref{f.gtilcs}, also
showing the results of uniform binning for reference. The plots also indicate
the ``good time intervals'' in which background flares were not an issue; they
are continuous except for the observations of \obj{g208-44-45}.
Table~\ref{t.xfluxes} includes information on the flare durations and fluxes.
Of the five sources with more than one block, four of them contain two blocks,
suggesting partially-observed flares. The last, \obj{lp851}, has three blocks,
with the pre- and post-flare fluxes agreeing at the \apx20\% level. From
visual inspection of Figure~\ref{f.gtilcs}, one may conclude that both a
flatter prior on $N_b$ (i.e., an assumption of a higher likelihood that
sources are variable) and that a non-piecewise-constant emission model would
lead to more faithful approximations of the data. Because the aim of our
analysis is limited to identifying representative quiescent and (when
appropriate) flaring X-ray fluxes, we do not explore these possible
elaborations here.

\begin{deluxetable*}{cllccr@{}l@{\,}l}

\tablecolumns{8}
\tablewidth{0em}
\tablecaption{Results of \chandra\ Analysis\label{t.xfluxes}}
\tablehead{
\colhead{~~} & \colhead{Name} & \colhead{State} & \colhead{Integ. Time} & \colhead{Counts} & \multicolumn{3}{c}{[$f_X$]} \\
 &  &  & \colhead{(s)} &  & \multicolumn{3}{c}{[\cgsflux]} \\ \\
\multicolumn{1}{c}{} & \multicolumn{1}{c}{(1)} & \multicolumn{1}{c}{(2)} & \multicolumn{1}{c}{(3)} & \multicolumn{1}{c}{(4)} & \multicolumn{3}{c}{(5)}
}
\startdata
\sidehead{\textit{Newly-observed targets}} & \obj{lhs292} & M & $10620$ & $51$ & $-13$ & $.7$  & $\pm\,0.1$\rule{0pt}{3ex} \\
 & \obj{lhs523} & M & $19605$ & $15$ & $-14$ & $.5$  & $\pm\,0.1$\rule{0pt}{3ex} \\
 & \obj{lhs2397} & M & $19804$ & $252$ & $-13$ & $.29$  & $\pm\,0.04$\rule{0pt}{3ex} \\
 & \obj{lhs3406} & M & $19804$ & $74$ & $-13$ & $.8$  & $\pm\,0.1$\rule{0pt}{3ex} \\
 &  & Q & $17342$ & $56$ & $-13$ & $.9$  & $\pm\,0.1$\rule{0pt}{3ex} \\
 &  & F & $2462$ & $18$ & $-13$ & $.3$  & $\pm\,0.1$\rule{0pt}{3ex} \\
 & \obj{lp647} & M & $19798$ & $12$ & $-14$ & $.6$  & $\pm\,0.2$\rule{0pt}{3ex} \\
 &  & Q & $14504$ & $3$ & $-15$ & $.0$  & $\pm\,0.2$\rule{0pt}{3ex} \\
 &  & F & $5294$ & $9$ & $-14$ & $.1$  & $\pm\,0.2$\rule{0pt}{3ex} \\
 & \obj{lp851} & M & $19800$ & $43$ & $-14$ & $.0$  & $\pm\,0.1$\rule{0pt}{3ex} \\
 &  & Q & $18126$ & $26$ & $-14$ & $.2$  & $\pm\,0.1$\rule{0pt}{3ex} \\
 &  & F & $1675$ & $17$ & $-13$ & $.3$  & $\pm\,0.1$\rule{0pt}{3ex} \\
 & \obj{n40026} & M & $18812$ & $6$ & $-14$ & $.8$  & $\pm\,0.2$\rule{0pt}{3ex} \\
\sidehead{\textit{\chandra\ archival targets}} & \obj{g208-44} & M & $12089$ & $234$ & $-13$ & $.05$  & $\pm\,^{0.04}_{0.05}$\rule{0pt}{3ex} \\
 & \obj{g208-45} & M & $12089$ & $170$ & $-13$ & $.22$  & $\pm\,^{0.07}_{0.08}$\rule{0pt}{3ex} \\
 &  & Q & $7459$ & $60$ & $-13$ & $.4$  & $\pm\,0.1$\rule{0pt}{3ex} \\
 &  & F & $4630$ & $110$ & $-13$ & $.0$  & $\pm\,0.1$\rule{0pt}{3ex} \\
 & \obj{dp0255} & M & $27908$ & $1$ & \multicolumn{3}{c}{$<$$-15.2$}\rule{0pt}{3ex} \\
 & \obj{lp349} & M & $37197$ & $60$ & $-14$ & $.1$  & $\pm\,0.1$\rule{0pt}{3ex} \\
 &  & Q & $9926$ & $6$ & $-14$ & $.5$  & $\pm\,0.2$\rule{0pt}{3ex} \\
 &  & F & $27270$ & $54$ & $-14$ & $.0$  & $\pm\,0.1$\rule{0pt}{3ex}
\enddata
\tablecomments{Col. (2) is the source state, one of mean (M), quiescent (Q), or flaring
(F). Col. (3) is the integration time after background flares have been
removed. Col. (4) is counts in the final apertures, accounting for the
flagging of periods of background flaring and removal of piled-up pixels. Col.
(5) is the X-ray flux in the 0.2--2~keV band.}
\end{deluxetable*}

We determined X-ray fluxes in the 0.2--2~keV band using either spectral
modeling or a simple energy conversion factor (ECF). We used the ECF approach
for all but the three brightest sources: \obj{lhs2397}, \obj{g208-44}, and
\obj{g208-45}. After extracting spectra from the event data of these three
sources, grouping into bins of $\ge$12 events, we used \sherpa\ version~1
\citep{thesherpa} for the modeling, ignoring energies outside of the standard
ACIS energy filter of 0.3--7~keV. We used the \sherpa\ implementation of the
\citet{thesimplex} simplex algorithm to optimize the modified $\chi^2$
statistic of \citet{g86}. Solar abundances were taken from \citet{l03}. Our
X-ray flux results and the outcomes of the modeling are discussed below. We
summarize the fluxes in Table~\ref{t.xfluxes}, where the uncertainties are
properly propagated and account for Poisson statistics as appropriate. Derived
quantities for the UCDs, including X-ray luminosities, are presented in
Table~\ref{t.dbdata}.

\subsection{\objn{lhs2397}}

\begin{figure}[htbp]
  \plotone{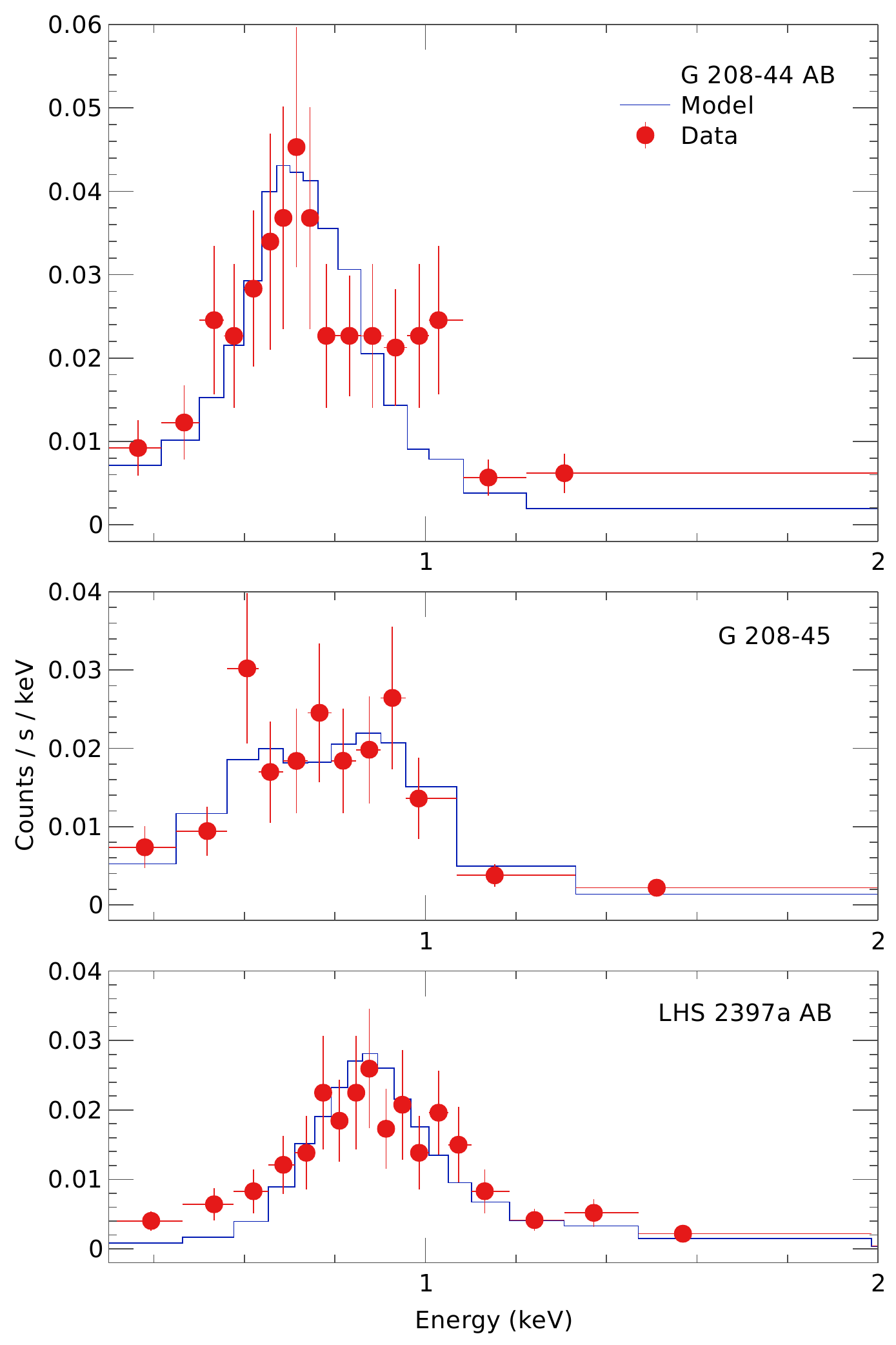}
  \caption{Mean X-ray spectra of \obj{g208-44}, \obj{g208-45}, and
    \obj{lhs2397} (\textit{red}) and best-fit spectral models (\textit{blue}).
    The horizontal lines associated with each data point indicate bin widths.
    \obj{g208-44} is modeled as a 0.30~keV plasma, \obj{g208-45} as
    two-temperature plasma with $kT_1 = 0.22$~keV and $kT_2 = 0.87$~keV, and
    \obj{lhs2397} as a 0.78~keV plasma. See text for caveats on the details of
    the model results.}
  \label{f.xspecs}
\end{figure}

In \obj{lhs2397}, a one-temperature, solar-abundance APEC \citep[Astrophysical
  Plasma Emission Code;][]{theapec} model yields a satisfactory fit, achieving
a reduced statistic $\chi_r^2=1.1$ with 18 degrees of freedom (DOF), although
there are strong correlations among the residuals. The data and best-fit model
are shown in Figure~\ref{f.xspecs}.

The model-fitting procedure finds $kT = 0.78 \pm 0.04$~keV, but we caution
generally against overinterpreting the parameters derived from our spectral
fits. While the data we present cannot rule out a single-temperature
solar-abundance model, the truth is likely more complicated. In particular,
the value of $kT$ we report above is quite possibly an approximate average of
multiple temperature components. Intriguingly, if we overfit the data by
adopting a two-temperature model, we find temperatures of \apx0.3 and
\apx1.2~keV, in good agreement with other results from mid-to-late M~dwarfs
\citeeg{rs05}. Similarly, if we use a single-temperature model with variable
abundances, the fits tend to converge on an inverse first ionization potential
(FIP) effect, with higher abundances found for elements with higher FIPs, as
is commonly seen in higher-S/N spectra of similar objects \citeeg{rs05}. We
emphasize, however, that these findings are not statistically significant.

Granting these caveats, the fluxes we derive from our spectral modeling are
robust, because the essential shapes of the X-ray spectra are well-constrained
by the data. For \obj{lhs2397} we find the X-ray flux in the 0.2--2~keV band
to be $\fx = -13.29\pm0.04$. At our adopted distance, this corresponds to $\lx
= 27.1$ and $\lxlb = -3.1$, near the canonical ``saturation'' value of the
X-ray/rotation activity relation observed in solar-type stars \citeeg{pmm+03}.

\subsection{\objn{g208-44-45}}
\label{s.g208}

While \obj{g208-44} and \obj{g208-45} are clearly resolved in the
\chandra\ image (Figure~\ref{f.g208img}), the tighter binary
(\objt{g208-44}\,A and B) is not. The following analysis considers only the
blended emission of \obj{g208-44}.

During our analysis of the \obj{g208-44-45} dataset, we discovered significant
background flaring activity. We used standard CIAO tools to extract a
light curve of non-source regions on the source chip (S3), binning by 150~s in
time and finding a non-flaring background rate of 0.32~s$^{-1}$ across the
whole chip, which is consistent with typical nonflaring behavior. We flagged
bins in which the measured background rate varied from this value by
$>$4$\sigma$. After processing, the good exposure time was reduced from 23.8
to 12.1~ks. We note that the Bayesian blocks method can be applied to data
with observational gaps without adjustment \citep{snjc13}, making it
well-suited for these data.

Both components of the system additionally achieved count rates sufficiently
high ($>$0.1~s$^{-1}$) to make pileup a concern. We used the Portable,
Interactive Multi-Mission Simulator (\pimms) to estimate the pileup
percentages during the observations. For both objects, pileup was estimated at
the \apx5\% level. To compensate for this we analyzed both components using
annuli centered on the source positions, removing the centermost pixels where
the count rate and hence pileup were greatest. For both sources, the annuli
had inner (outer) radii of 0.5 (2.25) arcsec. This approach discards
significant signal, but both sources were bright enough that we still retained
strong detections sufficient for spectral modeling. The reported fluxes
account for the reduced portion of the PSF being sampled. Filtering of
background flares and removal of the central pixels reduced the number of
counts detected at the position of \obj{g208-44} from 1607 to 234; for
\obj{g208-45}, the numbers are 1072 and 170, respectively.

In \obj{g208-44}, the Bayesian blocks analysis detects no significant
variability, and a one-temperature, solar-abundance APEC model yields a
satisfactory fit, achieving $\chi_r^2=1.2$ with 16 DOF. In this case as well
there are strong correlations among the residuals. The best-fit temperature in
the adopted model is $kT=0.30\pm0.02$~keV, subject to the same caveats
mentioned above. We find $\fx = -13.05^{+0.04}_{-0.05}$ (0.2--2~keV). At our
adopted distance, $\lx = 26.4$ and $\lxlb = -4.49$.

In \obj{g208-45}, the Bayesian blocks analysis finds that the count rate in
the final 9~ks of the observation is elevated by a factor of \apx3. Although
flaring is often associated with spectral variability \citeeg{oha+05}, there
were few enough counts available that we chose to only model the mean spectrum
of the source. We found that a one-temperature, solar-abundance model yielded
a less satisfactory fit, with $\chi_r^2 = 2.0$ for 11~DOF. A two-temperature,
solar-abundance model yields $\chi_r^2 = 0.8$ (9~DOF) with
$kT_1=0.22\pm0.5$~keV and $kT_2=0.87\pm0.14$~keV. We find
$\fx=-13.22^{+0.07}_{-0.08}$ (0.2--2~keV), $\lx = 26.2$, and $\lxlb=-4.48$.
For comparison, a one-temperature fit with variable abundance finds $\chi^2_r
= 1.2$ (10 DOF), $kT=0.59^{+0.10}_{-0.15}$~keV,
$Z/Z_\odot=0.08^{+0.05}_{-0.03}$, and $\fx=-13.35\pm0.35$. Despite their
differences, the two models yield values of \Fx\ that agree within their
uncertainties.

\subsubsection{Did ROSAT observe \objn{g208-44-45} during a flare?}

Previous X-ray observations of the \obj{g208-44-45} system were performed as
part of the \textit{ROSAT} All-Sky Survey (RASS), with follow-up on the
\textit{ROSAT} High Resolution Imager (HRI). While the latter was capable of
resolving the triple system into its two main components (resolution
\apx2$''$; separation \apx7$''$), the former, with a resolution of
\apx5--10$'$ \citep{therassbsc}, was not. While \citet{sl04} correctly
identify the ROSAT source as the blend of all three components, we note that
\citet{lhm08} failed to highlight this in their summary of the X-ray
properties of \obj{g208-44}.

The RASS catalog luminosity for \obj{g208-44-45} is $\lx=27.47$ in the
0.1--2.4~keV band. After correction to the ROSAT bandpass, this exceeds the
sum of our resolved measurements by a factor of \apx6. This suggests that the
RASS observations of \obj{g208-44-45} may have occurred during a strong flare.
The RASS exposure time during the observation was 730~s \citep{sl04}, shorter
than the typical timescale of such events. The later \textit{ROSAT} HRI
measurements find $\lx = 27.19$ with an integration time of 2849 seconds
\citep{sl04}, in better agreement with the \chandra\ results, though still
exceeding them by a factor of \apx2. As X-ray flares may easily result in
luminosity increases of an order of magnitude (cf.~Figure~\ref{f.lxspt}), it
would not take an unusually large flare to reconcile the two measurements. On
the other hand, the fact that both \textit{ROSAT} measurements exceed the
combined \chandra\ flaring luminosity, despite their separation in time
(8~months), suggests that perhaps the X-ray activity of this system has
decreased since the time of the RASS, possibly due to a long-term magnetic
activity cycle \citeeg{bnrss96}.

\subsection{Other Detected Sources}

We determined X-ray fluxes for the other detected sources by assuming an
energy conversion factor (ECF) of $(4.5\pm1)\times10^{-12}$
erg~cm$^{-2}$~count$^{-1}$, where the uncertainty approximately accounts for
the range of plasma temperatures and abundances commonly encountered. The
applicable theoretical ECF reported by \textsf{WebPIMMS} version 4.6a for a
0.5~keV APEC plasma with $Z/Z_\odot = 0.6$ is $4.4\times10^{-12}$
erg~cm$^{-2}$~count$^{-1}$. Our adopted value agrees with those derived for
\obj{lhs2397}, \obj{g208-44}, and \obj{g208-45} (ECF $= (4.3, 4.6,
4.3)\times10^{-12}$, respectively) as well as our previous observations of
X-ray-emitting UCDs \citep{bgg+08,bbg+08}. Spectral modeling of the other
sources with larger numbers of events (\obj{lhs292}, \obj{lhs3406},
\obj{lp851}) yields results consistent with those reported here.

\subsection{Undetected Source: \objn{dp0255}}

Only one photon in the 0.3--7~keV range was detected at the predicted
location of \obj{dp0255} throughout both observations, compared to the
expected background level of 0.7~counts in 28~ks. The resulting 95\%
confidence upper limit is 4.3 counts \citep{kbn91}. Using the above energy
conversion factor, the time-averaged flux limit is $\fx < -15.2$.
At our adopted distance of 5.0~pc, we can thus constrain the persistent
emission of \obj{dp0255} to be $\lx \lesssim 24.3$, or $\lxlb\lesssim-4.70$.

\section{Radio Analysis}
\label{s.radio}

We calibrated the VLA using standard procedures in the CASA software system
\citep{thecasa}. Radio-frequency interference was flagged automatically using
the \textsf{aoflagger} tool, which provides post-correlation \citep{odbb+10}
and morphological \citep{ovdgr12} algorithms for identifying interference. At
the time of analysis, \textsf{aoflagger} did not include a set of tuning
parameters suitable for the processing of VLA data, so these were developed
manually.

\begin{deluxetable*}{lcccccc}

\tablecolumns{7}
\tablewidth{0em}
\tablecaption{Results of VLA Analysis\label{t.rfluxes}}
\tablehead{
\colhead{Name} & \colhead{Integ. Time} & \colhead{$S_\nu$} & \colhead{rms} & \multicolumn{3}{c}{Synthesized Beam}   \\
 \cline{5-7}  & \colhead{} & \colhead{} & \colhead{} & \colhead{Major} & \colhead{Minor} & \colhead{PA} \\
 & \colhead{(s)} & \colhead{(\ujy)} & \colhead{(\ujybm)} & \colhead{(arcsec)} & \colhead{(arcsec)} & \colhead{(deg)} \\ \\
\multicolumn{1}{c}{(1)} & \multicolumn{1}{c}{(2)} & \multicolumn{1}{c}{(3)} & \multicolumn{1}{c}{(4)} & \multicolumn{1}{c}{(5)} & \multicolumn{1}{c}{(6)} & \multicolumn{1}{c}{(7)}
}
\startdata
\obj{lhs292} & $2260$ & <$23$ & $7.7$ & $5.0$ & $2.7$ & $-32$ \\
\obj{lhs523} & $4405$ & <$13$ & $4.4$ & $4.2$ & $2.5$ & $41$ \\
\obj{lhs2397} & $2335$ & $63 \pm 7$ & $5.3$ & $4.2$ & $2.8$ & $1$ \\
\obj{lhs3406} & $2205$ & <$16$ & $5.4$ & $3.5$ & $3.0$ & $108$ \\
\obj{lp647} & $2265$ & <$22$ & $7.4$ & $4.5$ & $2.9$ & $33$ \\
\obj{lp851} & $2205$ & <$19$ & $6.2$ & $7.1$ & $2.8$ & $-29$ \\
\obj{n40026} & $2205$ & <$16$ & $5.3$ & $3.1$ & $2.7$ & $-43$
\enddata
\tablecomments{Col. (4) is the background rms in a region near the source. Cols. (5) and
(6) are FWHM sizes. The reference frequency of each image is 6.05~GHz. Further
parameters regarding the detection of \obj{lhs2397} are given in the text.}
\end{deluxetable*}

We created deep Stokes~I images of each field with 2048$\times$2048 pixels,
each $1$$\times$$1$ arcsec$^2$, except the image of \obj{lhs523}, for which a
pixel scale of $1.5$$\times$$1.5$ arcsec$^2$ was used to include the nearby,
bright blazar \objectname[QSO B2227-136]{QSO\,B2227$-$136} in the image. The
imaging process used multi-frequency synthesis \citep{themfs} and CASA's
multi-frequency CLEAN algorithm with 1500 iterations. Two spectral Taylor
series terms were used for each CLEAN component; this approach models both the
flux and spectral index of each source. The reference frequency for each image
is 6.05~GHz. Properties of the images are listed in Table~\ref{t.rfluxes}.
Astrometric predictions of the source locations were computed as described
above, and cutouts of the VLA images around the predicted source locations are
show in Figure~\ref{f.cutouts}. The accuracy of VLA astrometry in our
observing configuration is \apx$1''$, comparable to that of our predictions.
While our targets have relatively high proper motions, the time baseline
between the pairs of VLA and \chandra\ observations is sufficiently small that
the differences in the predicted positions are negligible.

We detect a radio source in the image of the \obj{lhs2397} field at position
RA = 11:21:48.78, Dec = $-$13:13:09.4, coincident with our astrometric
prediction of RA = 11:21:48.77, Dec = $-$13:13:09.5. We re-imaged this field,
rephasing the data to place this position on a pixel center to obtain the most
accurate source parameters from image-domain modeling. Fitting the rephased
image with a point-source model yields a flux density of $63\pm7$~\ujy\ and a
positional uncertainty of $0.4''$. \citet{fwkk91} used the VLA to find the
areal density of sources brighter than 50~\ujy\ at 5~GHz to be \apx$6.4 \times
10^{-5}$ arcsec$^{-2}$, making the probability of a chance positional
coincidence \apx$2 \times 10^{-3}$ (taking our search area to be the
synthesized beam size). We therefore identify this source with \obj{lhs2397}.
The multi-frequency cleaning algorithm determines a spectral index of $\alpha =
-0.4 \pm 0.2$ ($S_\nu \propto \nu^\alpha$). We imaged the source in the
Stokes~V parameter and made no detection, with an image rms of 5.4~\ujybm.
Taking the Stokes~V upper limit to be three times this value, we find that
$|V|/I \lesssim 25\%$. We searched for flares and other forms of variability
using a visibility-domain analysis of the Stokes~I data as described in
\citet{wbz13}. No significant indications of variability were seen.

There is a 50~\ujy\ (7$\sigma$) radio source in the image of the \obj{lhs3406}
field at RA = 18:43:20.72, Dec = $+$40:40:33.01, which is 15$''$ distant from
the astrometric prediction. We checked the image astrometry against two NVSS
sources (\object{NVSS\,J184331$+$404756} and \object{NVSS\,J184314$+$403302}),
finding agreement down to the \apx$1.5''$ uncertainty in the survey's
astrometry. We conclude that this source is not \obj{lhs3406}. There are no
sources within \apx1$'$ of our astrometric predictions in all of the other
fields. In each of these cases, we place an upper limit on the target flux
density of three times the image rms. The field of \obj{lhs523} was visited
twice; no source is detected in the individual visits or in a deep image
formed by combining the two datasets. The results of our radio observations,
including flux densities, are summarized in Table~\ref{t.rfluxes}. Derived
parameters, including radio spectral luminosities, are presented in
Table~\ref{t.dbdata}.

\section{Trends in Radio and X-Ray Emission}
\label{s.trends}

We have combined our new measurements with data from the literature to compile
a comprehensive database of UCDs with both radio and X-ray observations. In
Table~\ref{t.dbsrcs}, we list these objects and provide some of their properties.
Different authors report X-ray luminosities that are integrated over varying
energy regimes; for consistency we normalize all X-ray fluxes and luminosities
to a common band of 0.2--2.0 keV. We used \pimms\ to compute the appropriate
conversion factors, evaluating flux ratios in a range of different plasma
temperatures in the APEC model. The resulting factors are listed in
Table~\ref{t.xbconvs}. The conversion factors represent an approximate median
for several temperatures in the range $kT=0.4$--$1.0$~keV and are stable to
within $5\%$ for temperatures within this range. In Table~\ref{t.dbdata} we
report all paired UCD radio and X-ray luminosities available, giving detailed
references and using simultaneous measurements when available.

\begin{deluxetable*}{r@{\,}lllr@{}lr@{}lr@{}lr@{}lcc}

\tablecolumns{14}
\tablewidth{0em}
\tablecaption{UCDs with Both Radio and X-ray Measurements\label{t.dbsrcs}}
\tablehead{
\colhead{} & \colhead{2MASS Identifier} & \colhead{Other Name} & \colhead{SpT} & \multicolumn{2}{c}{$J$} & \multicolumn{2}{c}{$K$} & \multicolumn{2}{c}{$d$} & \multicolumn{2}{c}{[$\Lb$]} & \multicolumn{2}{c}{References}  \\
 \cline{13-14}  &  &  &  & \multicolumn{2}{c}{(mag)} & \multicolumn{2}{c}{(mag)} & \multicolumn{2}{c}{(pc)} & \multicolumn{2}{c}{[$L_\sun$]} &  &  \\ \\
\multicolumn{1}{c}{} & \multicolumn{1}{c}{(1)} & \multicolumn{1}{c}{(2)} & \multicolumn{1}{c}{(3)} & \multicolumn{2}{c}{(4)} & \multicolumn{2}{c}{(5)} & \multicolumn{2}{c}{(6)} & \multicolumn{2}{c}{(7)} & \multicolumn{1}{c}{(S)} & \multicolumn{1}{c}{(D)}
}
\startdata
\P & \object{10481258$-$1120082} & LHS 292 & M6.5 & $8$ & $.86$ & $7$ & $.93$ & $4$ & $.5$ & $-3$ & $.15$ & 1 & 2 \\
\P & \object{22285440$-$1325178} & GJ 4281 & M6.5 & $10$ & $.77$ & $9$ & $.84$ & $11$ & $.3$ & $-3$ & $.13$ & 3 & 2 \\
 & \object{13142039$+$1320011 AB} & NLTT 33370 AB & M7 & $9$ & $.75$ & $8$ & $.79$ & $16$ & $.4$ & $-2$ & $.40$ & 4 & 4 \\
 & \object{14563831$-$2809473} & LHS 3003 & M7 & $9$ & $.96$ & $8$ & $.93$ & $6$ & $.4$ & $-3$ & $.29$ & 5 & 2 \\
 & \object{16553529$-$0823401} & vB 8 & M7 & $9$ & $.78$ & $8$ & $.82$ & $6$ & $.5$ & $-3$ & $.21$ & 1 & 6 \\
\P & \object{11554286$-$2224586} & LP 851$-$346 & M7.5 & $10$ & $.93$ & $9$ & $.88$ & $9$ & $.7$ & $-3$ & $.32$ & 7 & 7 \\
\P & \object{15210103$+$5053230} & NLTT 40026 & M7.5 & $12$ & $.01$ & $10$ & $.92$ & $16$ & $.1$ & $-3$ & $.30$ & 5 & 5 \\
\P & \object{00275592$+$2219328 AB} & LP 349$-$25 AB & M8 & $10$ & $.61$ & $9$ & $.57$ & $13$ & $.2$ & $-2$ & $.93$ & 5 & 8 \\
 & \object{03205965$+$1854233} & LP 412$-$31 & M8 & $11$ & $.76$ & $10$ & $.64$ & $14$ & $.5$ & $-3$ & $.29$ & 5 & 9 \\
\P & \object{11214924$-$1313084 AB} & LHS 2397a AB & M8 & $11$ & $.93$ & $10$ & $.73$ & $14$ & $.3$ & $-3$ & $.36$ & 10 & 2 \\
\P & \object{18432213$+$4040209} & LHS 3406 & M8 & $11$ & $.31$ & $10$ & $.31$ & $14$ & $.1$ & $-3$ & $.16$ & 5 & 2 \\
 & \object{19165762$+$0509021} & vB 10 & M8 & $9$ & $.91$ & $8$ & $.77$ & $6$ & $.1$ & $-3$ & $.30$ & 11 & 11 \\
 & \object{14542923$+$1606039 Bab} & Gl 569 Bab & M8.5 & $10$ & $.61$ & $9$ & $.45$ & $9$ & $.8$ & $-3$ & $.17$ & 12 & 13 \\
 & \object{18353790$+$3259545} & LSPM J1835$+$3259 & M8.5 & $10$ & $.27$ & $9$ & $.17$ & $5$ & $.7$ & $-3$ & $.52$ & 5 & 9 \\
\P & \object{01095117$-$0343264} & LP 647$-$13 & M9 & $11$ & $.69$ & $10$ & $.43$ & $11$ & $.1$ & $-3$ & $.48$ & 5 & 5 \\
 & \object{03393521$-$3525440} & LP 944$-$20 & M9 & $10$ & $.72$ & $9$ & $.55$ & $5$ & $.0$ & $-3$ & $.81$ & 5 & 9 \\
 & \object{08533619$-$0329321} & LHS 2065 & M9 & $11$ & $.21$ & $9$ & $.94$ & $8$ & $.5$ & $-3$ & $.52$ & 5 & 2 \\
 & \object{10481463$-$3956062} &  & M9 & $9$ & $.54$ & $8$ & $.45$ & $4$ & $.0$ & $-3$ & $.54$ & 14 & 15 \\
 & \object{14284323$+$3310391} & LHS 2924 & M9 & $11$ & $.99$ & $10$ & $.74$ & $10$ & $.8$ & $-3$ & $.63$ & 15 & 2 \\
 & \object{15010818$+$2250020} & TVLM 513$-$46546 & M9 & $11$ & $.87$ & $10$ & $.71$ & $9$ & $.9$ & $-3$ & $.67$ & 15 & 2 \\
 & \object{00242463$-$0158201} & BRI B0021$-$0214 & M9.5 & $11$ & $.99$ & $10$ & $.54$ & $12$ & $.1$ & $-3$ & $.49$ & 16 & 2 \\
 & \object{00274197$+$0503417} & PC 0025$+$0447 & M9.5 & $16$ & $.19$ & $14$ & $.96$ & $72$ & $.0$ & $-3$ & $.67$ & 17 & 17 \\
 & \object{07464256$+$2000321 AB} &  & L0 & $11$ & $.76$ & $10$ & $.47$ & $12$ & $.2$ & $-3$ & $.36$ & 18 & 5 \\
 & \object{06023045$+$3910592} & LSR J0602$+$3910 & L1 & $12$ & $.30$ & $10$ & $.87$ & $10$ & $.6$ & $-3$ & $.67$ & 19 & 19 \\
 & \object{13054019$-$2541059 AB} & Kelu$-$1 AB & L2 & $13$ & $.41$ & $11$ & $.75$ & $18$ & $.7$ & $-3$ & $.56$ & 5 & 5 \\
 & \object{05233822$-$1403022} &  & L2.5 & $13$ & $.08$ & $11$ & $.64$ & $13$ & $.4$ & $-3$ & $.82$ & 5 & 9 \\
 & \object{00361617$+$1821104} & LSPM J0036$+$1821 & L3.5 & $12$ & $.47$ & $11$ & $.06$ & $8$ & $.8$ & $-3$ & $.99$ & 20 & 15 \\
 & \object{12281523$-$1547342 AB} &  & L5 & $14$ & $.38$ & $12$ & $.77$ & $20$ & $.2$ & $-3$ & $.93$ & 5 & 5 \\
 & \object{15074769$-$1627386} &  & L5 & $12$ & $.83$ & $11$ & $.31$ & $7$ & $.3$ & $-4$ & $.23$ & 20 & 9
\enddata
\tablecomments{Rows marked with a pilcrow (\P) indicate sources with new measurements
presented in this work. Col. (3) is spectral type. Cols. (4) and (5) are from
2MASS \citep{the2mass}. Col. (7) is the bolometric luminosity, the calculation
of which is described in the Appendix; note that here it is given in units of
$L_\odot$, not cgs.}
\tablerefs{Columns are (S), spectral type; and (D), distance. [1] \citet{hks94}, [2] \citet{thegctp}, [3] \citet{khm91}, [4] \citet{ltsr09}, [5] \citet{crl+03}, [6] \citet{thethirdgj}, [7] \citet{cpbd+05}, [8] \citet{gc09}, [9] \citet{crk+07}, [10] \citet{fcs03}, [11] \citet{bbg+08}, [12] \citet{zolp+04}, [13] \citet{s04}, [14] \citet{rb10}, [15] \citet{rck+08}, [16] \citet{rhg95}, [17] \citet{mbr12}, [18] \citet{brpb+09}, [19] \citet{bbf+10}, [20] \citet{brr+05}}
\end{deluxetable*}

\begin{deluxetable}{lc}

\tablecolumns{2}
\tablewidth{0em}
\tablecaption{X-Ray Band Conversion Factors\label{t.xbconvs}}
\tablehead{
\colhead{Band} & \colhead{Factor} \\
\colhead{(keV)} & 
}
\startdata
$0.1$--$2.4$ & $0.84$ \\
$0.1$--$10.0$ & $0.80$ \\
$0.2$--$8.0$ & $0.95$ \\
$0.3$--$0.8$ & $1.53$ \\
$0.3$--$2.0$ & $1.10$ \\
$0.3$--$7.0$ & $1.05$ \\
$0.3$--$8.0$ & $1.05$ \\
$0.3$--$10.0$ & $1.05$ \\
$0.5$--$8.0$ & $1.20$
\enddata
\end{deluxetable}

Although we focused on X-ray and radio fluxes when constructing our database,
it contains many ancillary measurements such as distances, spectral types,
photometry, and effective temperatures. It is compiled from simple textual
tables that are maintained in the Git distributed version control system,
taking inspiration from the architecture of the Open Exoplanet Catalogue
\citep{r12barxiv}. Its design is intended to enable continuous refinement in a
decentralized, collaborative manner. Further details will be presented in a
future publication.

\subsection{X-Ray Luminosity vs. Spectral Type}
\label{s.lxspt}

\begin{figure*}[htbp]
  \plotone{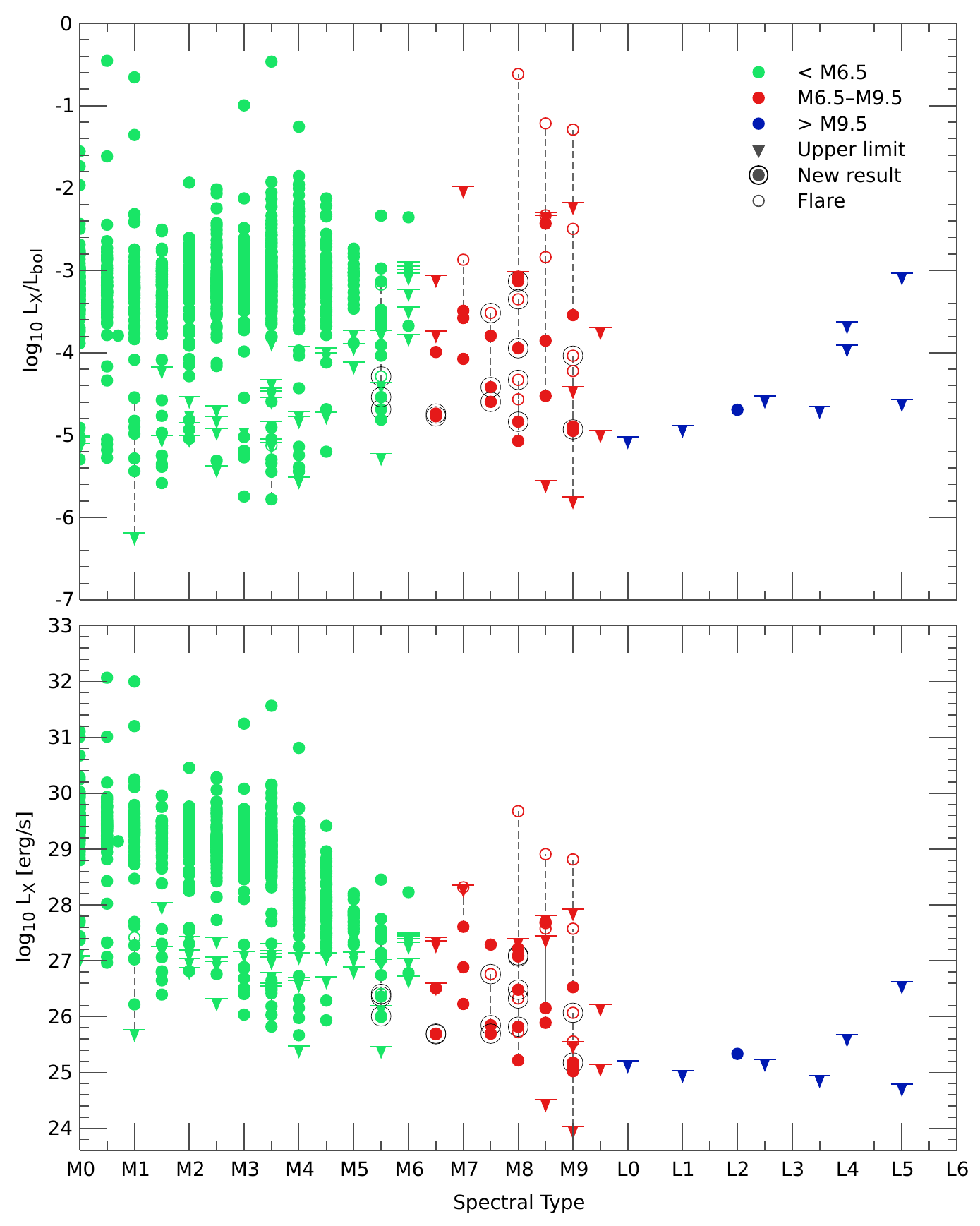}
  \caption{X-ray luminosity as a function of spectral type, both normalized by
    \Lb\ (\textit{upper panel}) and unnormalized (\textit{lower panel}).
    \Lx\ is in the 0.2--2~keV band. Objects with adopted spectral types of M6
    or earlier, M$6.5$--M$9.5$, and L0 or later are plotted in \textit{green},
    \textit{red}, and \textit{blue}, respectively. Upper limits are
    represented by \textit{downward-pointing triangles}. \textit{Dashed lines}
    connect multiple measurements of the same source, with \textit{open
      symbols} denoting known flaring emission. Measurements from this work
    are highlighted with \textit{black circles}. Both the normalized and
    unnormalized luminosities appear to have a typical scale in earlier-type
    objects ($10^{-3}$ and $10^{29.3}$~\ergps, respectively) with a falloff in
    later-type objects, although the starting point for the falloff differs
    between the two cases (\apx M5 and \apx M$3.5$, respectively) due to the
    evolution of \Lb\ with spectral type.}
  \label{f.lxspt}
\end{figure*}

Low-mass stars of spectral types earlier than M6 obey an X-ray
activity/rotation relationship with ``saturation'' at $\lxlb \sim -3$
\citep{pmm+03,jjb+11,wdmh11}. Because main-sequence late-type stars are generally
rapid rotators \citep{ib08}, most of them have X-ray emission around this
saturation level, as demonstrated in Figure~\ref{f.lxspt}. The $<$M6 objects
in this figure having $\lxlb \gtrsim -2$ are likely flares. All come from the
work of \citet{rgh06}, who obtained 1080 M~dwarf X-ray fluxes from the ROSAT
All-Sky Survey, a fraction of which will inevitably have been measured during
a flare. All of the $<$M6 objects with $\lxlb \lesssim -4.5$ and measured
\vsi\ are slow rotators, and thus their low levels of X-ray emission are
expected. There is a noticeable underdensity of sources with $\lxlb \sim -4$
or $\lx \sim 28$; it can also be seen clearly in the data of \citet{pmm+03}.
This underdensity is reminiscent of the ``Vaughan-Preston gap'' \citep{vp80},
a similar feature seen in the distribution of various chromospheric activity
indicators observed in F and G stars, for which a variety of explanations have
been offered, including changes in dynamo modes, evolutionary stages of rapid
angular momentum loss, or two distinct waves of star formation in the solar
neighborhood \citeeg{dmr81}. Somewhat surprisingly, we are unable to locate in
the literature any investigation of this feature; such an undertaking is
beyond the scope of this work.

Figure~\ref{f.lxspt} shows that our observations lend further support to the
conclusion that the standard X-ray ``saturation'' effect breaks down at
spectral types $\gtrsim$M6 \citep{fgg03,bbf+10}, an effect that is also seen
in \LhLb\ \citep{gmr+00,mb03}. Although the cause of this breakdown has been
the subject of much study, the number of detected objects is small while there
are many possibly-relevant physical effects: decreasing \teff, increasing
rotation, a disappearing radiative core, and an increasingly dipolar magnetic
field \citep{mdp+08}. The role of rotation is of particular interest because
of its known effect on X-ray activity and its strong correlation with spectral
type; this issue is explored in \papertwo.

The data show a similar breakdown in terms of \Lx, with $\lx \approx 29.3$ for
spectral types $\lesssim$M4 but decreasing by \apx1~dex for each later
spectral subtype. This empirical relationship indicates a breakdown of the
relationship around where full convection sets in, at spectral types of
$\gtrsim$M3.5--M4 \citep{cb00}. Furthermore, there are indications that it may
be more appropriate to consider \Lx\ rather than \LxLb\ in this regime:
results from Zeeman-Doppler imaging studies suggest that mid-to-late M~dwarfs
harbor relatively weak, disordered magnetic fields similar to those of
slowly-rotating solar-type stars \citep{mdp+10}, and there is substantially
less scatter in the slow-rotator region of the X-ray activity/rotation
relationship when \Lx\ rather then \LxLb\ is considered \citep{pmm+03}.

\subsection{The G\"udel-Benz Relation}

\begin{figure*}[htbp]
  \plotone{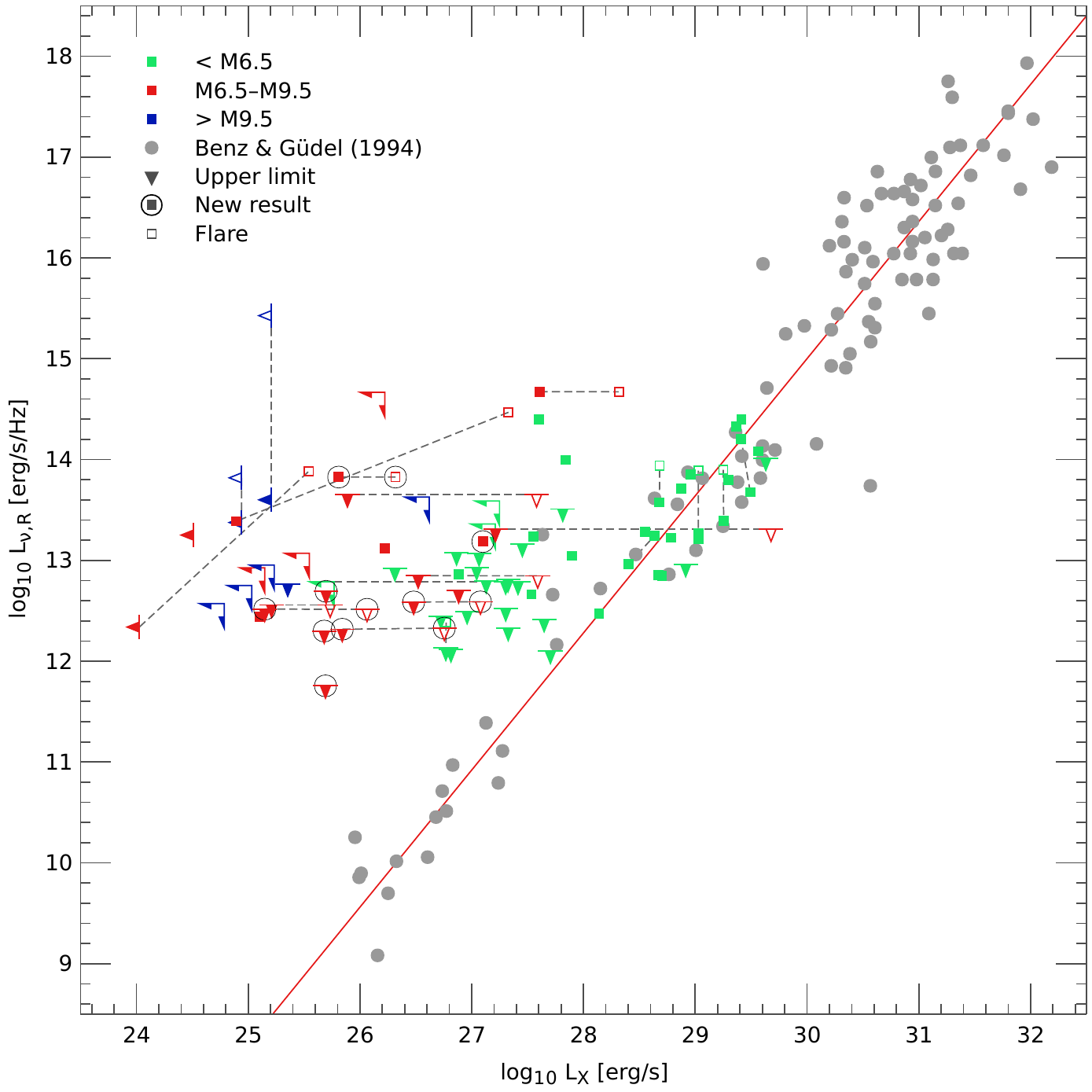}
  \caption{The ``G\"udel-Benz'' relationship (GBR) between \Lx\ (0.2--2~keV)
    and \sLr. Colors and symbols are as in Figure~\ref{f.lxspt}, with limits
    shown by triangles pointing down, left, or down-left. Data for objects
    M6.5 or later (\textit{red} and \textit{blue} symbols) are tabulated in
    Table~\ref{t.dbdata}. \textit{Gray circles} reproduce the original data of
    \citet{bg94}: those with $\slr < 12$ are solar flares; $12 < \slr \lesssim
    14.5$ are dMe and dKe stars; $\slr \gtrsim 14.5$ are active binaries.
    Except for \obj{lp349}, the new data do not show the extreme radio
    over-luminosity of some previous observations, being consistent with radio
    over-luminosities of $\lesssim$$10^2$ compared to the GBR.}
  \label{f.lxlr}
\end{figure*}

\begin{figure*}[htbp]
  \plotone{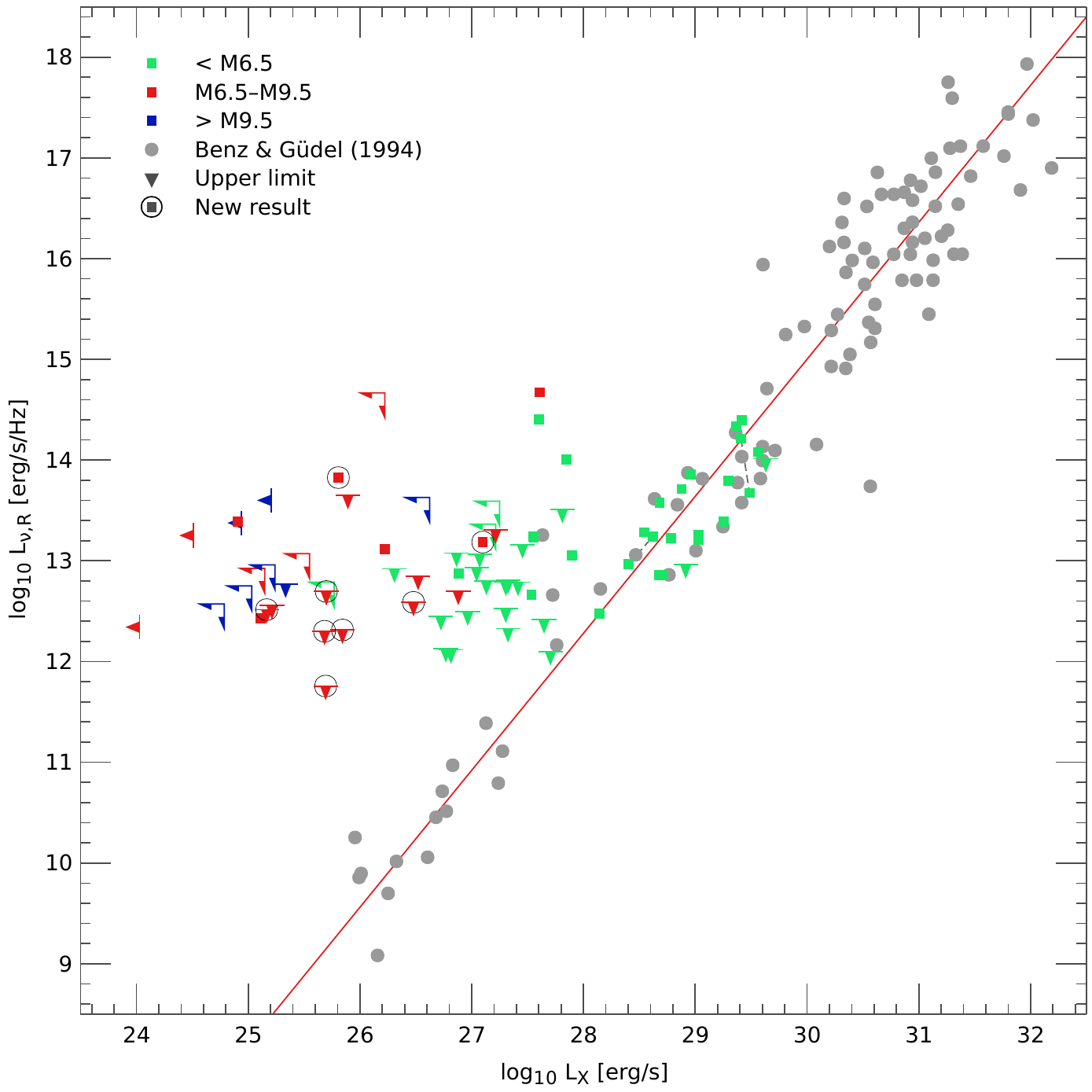}
  \caption{The same as Figure~\ref{f.lxlr}, but omitting known flares.}
  \label{f.lxlrnof}
\end{figure*}

Our paired radio and X-ray observations allow us to consider ultracool dwarfs
in the context of the GBR. This is particularly salient in the UCD regime
because the first detection of radio emission from a brown dwarf implied a
severe divergence from the GBR \citep{bbb+01}, and subsequent observations
have confirmed that this divergence is not uncommon \citeeg{bgg+08,bbf+10}. In
Figure~\ref{f.lxlr}, we plot our full database of UCDs with both radio and
X-ray measurements as well as the original data of \citet{bg94}.
Figure~\ref{f.lxlrnof} is similar but omits known flares. In these and several
subsequent plots, we show the best linear fit to the \citet{bg94} data from
\citet{bbf+10},
\begin{equation}
\slr = 1.36 (\lx - 18.97).
\label{e.gbfit}
\end{equation}
The scatter of the \citet{bg94} data around this fit is 0.6~dex when \Lx\ is
treated as an independent variable (i.e., the distance from the best-fit line
is measured at fixed \Lx). The scatter relative to the best-fit line (i.e.,
measured perpendicular to it) is 0.2~dex. In the \citet{bg94} dMe data,
$\slrlx \sim -15.5$ typically, a value we adopt as a reference when
quantifying radio over-luminosity relative to the GBR.

As may be seen in Figure~\ref{f.lxlr}, almost all of our new results could be
consistent with the established GBR. Since the radio upper limits are
\apx2~dex above the best-fit line. The lone new radio detection,
\obj{lhs2397}, has $\slrlx = -13.9$ and lies 1.3~dex away from the best-fit
line. It is thus an outlier but not nearly as extreme as, for example,
\obj{tvlm513}, which has $\slrlx = -11.5$ and is 3.2~dex away from the fit in
quiescence. We find no significant signs of variability in the emission of
\obj{lhs2397} in either band, which suggests that the observed fluxes
correspond to a quiescent rather than flaring state. \obj{lp349}, for which we
have combined a new X-ray analysis with a radio measurement from the
literature, has $\slrlx = -12.0$ and lies 2.7~dex away from the best-fit line
in quiescence, making it a strong violator of the GBR.

\begin{figure*}[htbp]
  \plotone{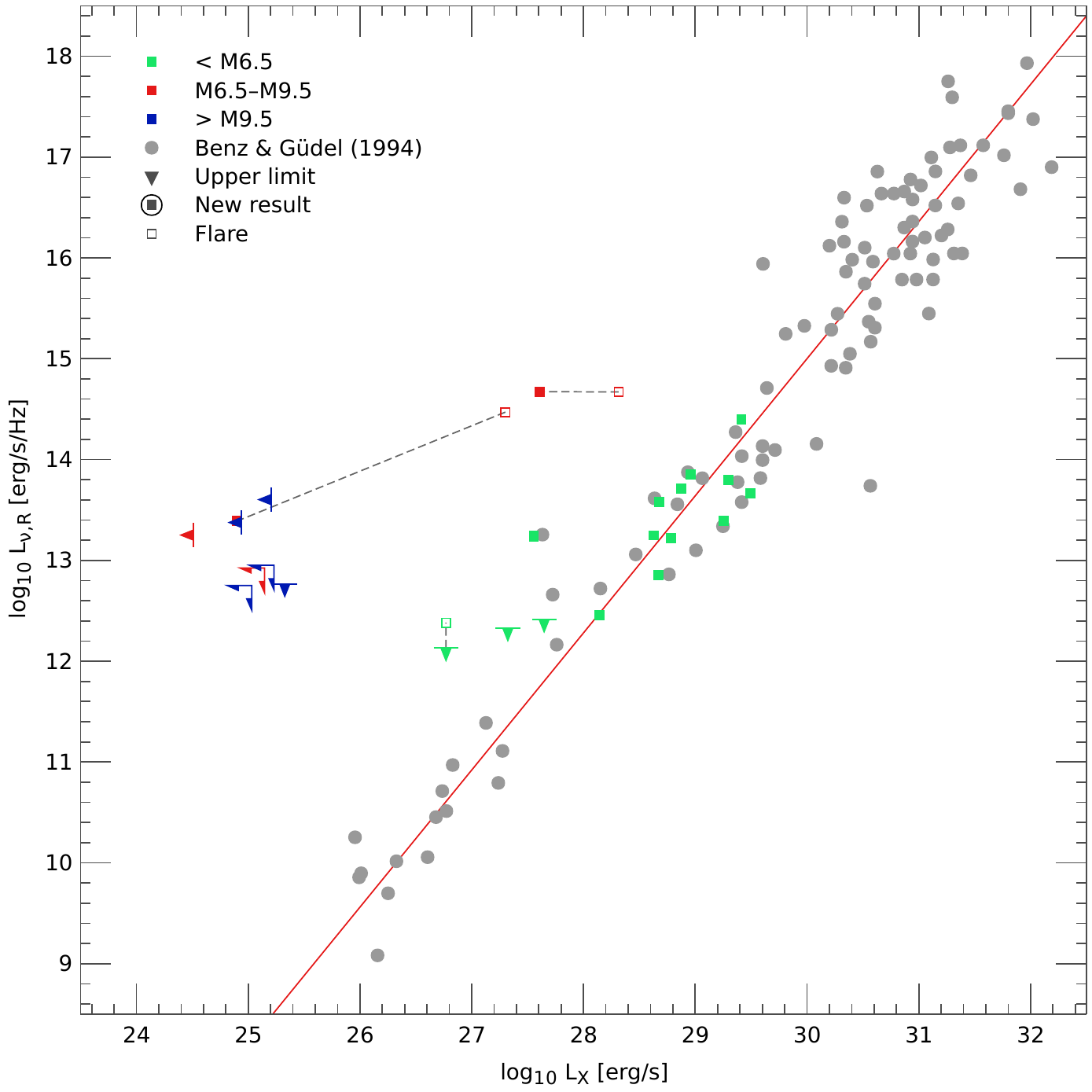}
  \caption{The same as Figure~\ref{f.lxspt}, but including only UCDs with
    simultaneous radio and X-ray observations. Some of the most extreme
    instances of UCD radio over-luminosity come from simultaneous observations,
    firmly establishing that variability is not responsible for the observed
    behavior. Objects earlier than M6.5 with simultaneous observations are
    consistent with the GBR, while those with non-simultaneous observations
    present a more ambiguous picture (cf.~Fig~\ref{f.lxlr}).}
  \label{f.lxlrsim}
\end{figure*}

UCD emission in both the radio and X-ray bands can show significant flaring
during typical observational timescales \citeeg{bbb+01,bp05,rbmb00,ssml06},
and the radio emission is additionally known to evolve over \apx year
timescales in some cases \citeeg{adh+07}. The simultaneity of measurements is
thus important to consider in the context of the GBR. In Table~\ref{t.dbdata},
we annotate which radio/X-ray measurements arise from simultaneous
observations, and we isolate these measurements in Figure~\ref{f.lxlrsim}.
Simultaneous observations constitute about one third of the existing dataset,
with many of the measurements coming from an observational campaign we have
conducted over the past several years \citep[Williams et al., in
  preparation]{brr+05, bbg+08, bgg+08, brpb+09, bbf+10}. These simultaneous
measurements include several of the most extreme GBR violators, showing that
strong divergence from the GBR is a genuine phenomenon and \textit{not} merely
due to flaring. There is only one simultaneously-observed UCD to be detected
in the X-ray but not the radio: the L2+L3.5 ($\pm$1 subtype) binary
\obj{kelu1} \citep{aob+07}. Its detection by \chandra\ was marginal (4
counts), precluding a detailed analysis.

\begin{figure}[htbp]
  \plotone{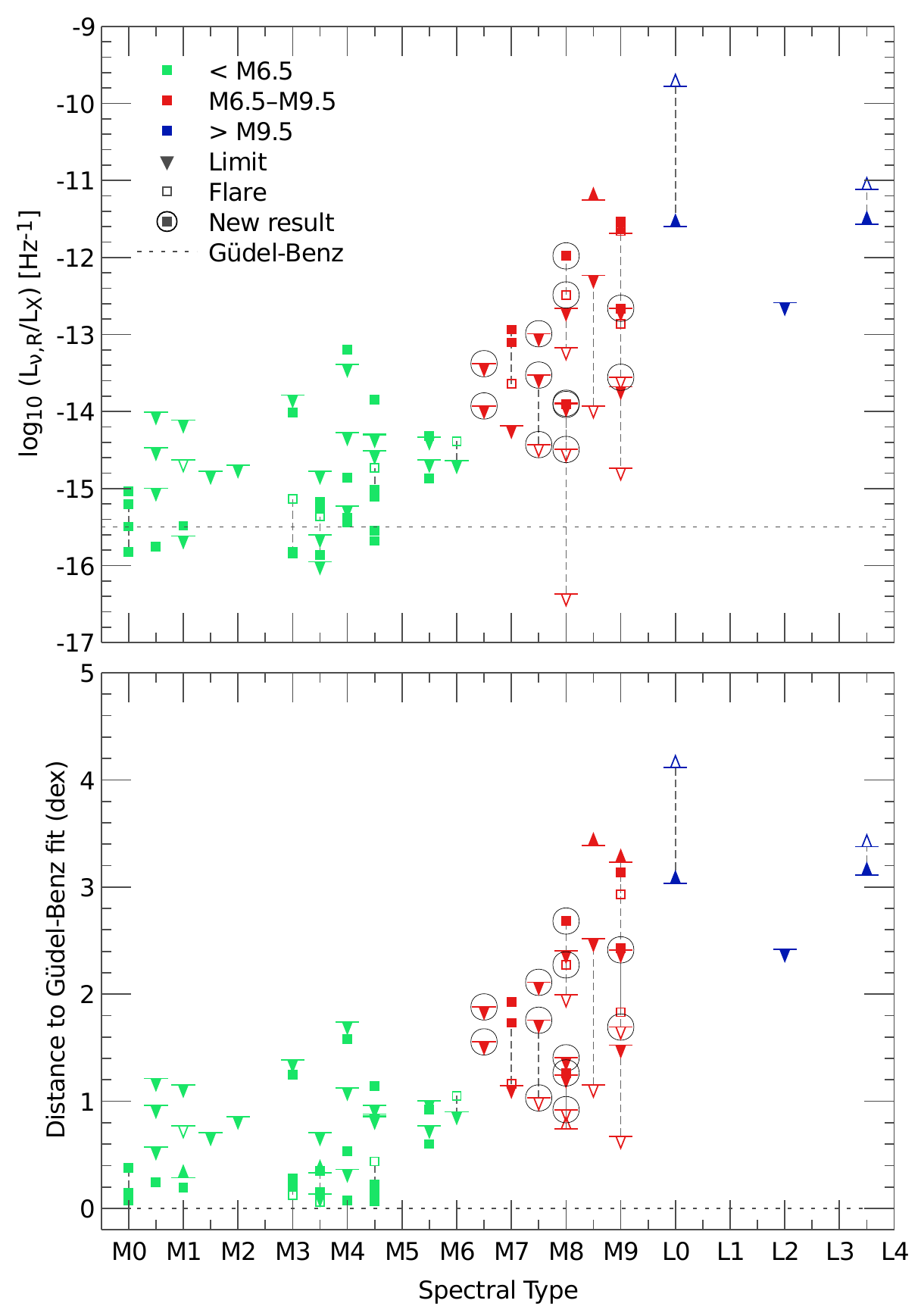}
  \caption{\textit{Upper panel:} measured ratios of radio to X-ray
    luminosities as a function of spectral type. Colors and symbols are as in
    Figure~\ref{f.lxspt}. The first measurement in which $[\sLr/\Lx] > -13$ is
    at spectral type M7. The reference line of $[\sLr/\Lx] = -15.5$ is
    appropriate for the GBR at SpT \apx\ M0--M6 \citep{bbf+10}. \textit{Lower
      panel:} same underlying data as the upper panel, but now plotting
    distance from the linear fit to the G\"udel-Benz relationship
    (Equation~\ref{e.gbfit}) rather than \sLrLx. The overall structure is
    virtually identical.}
  \label{f.rxspt}
\end{figure}

\begin{figure}[htbp]
  \plotone{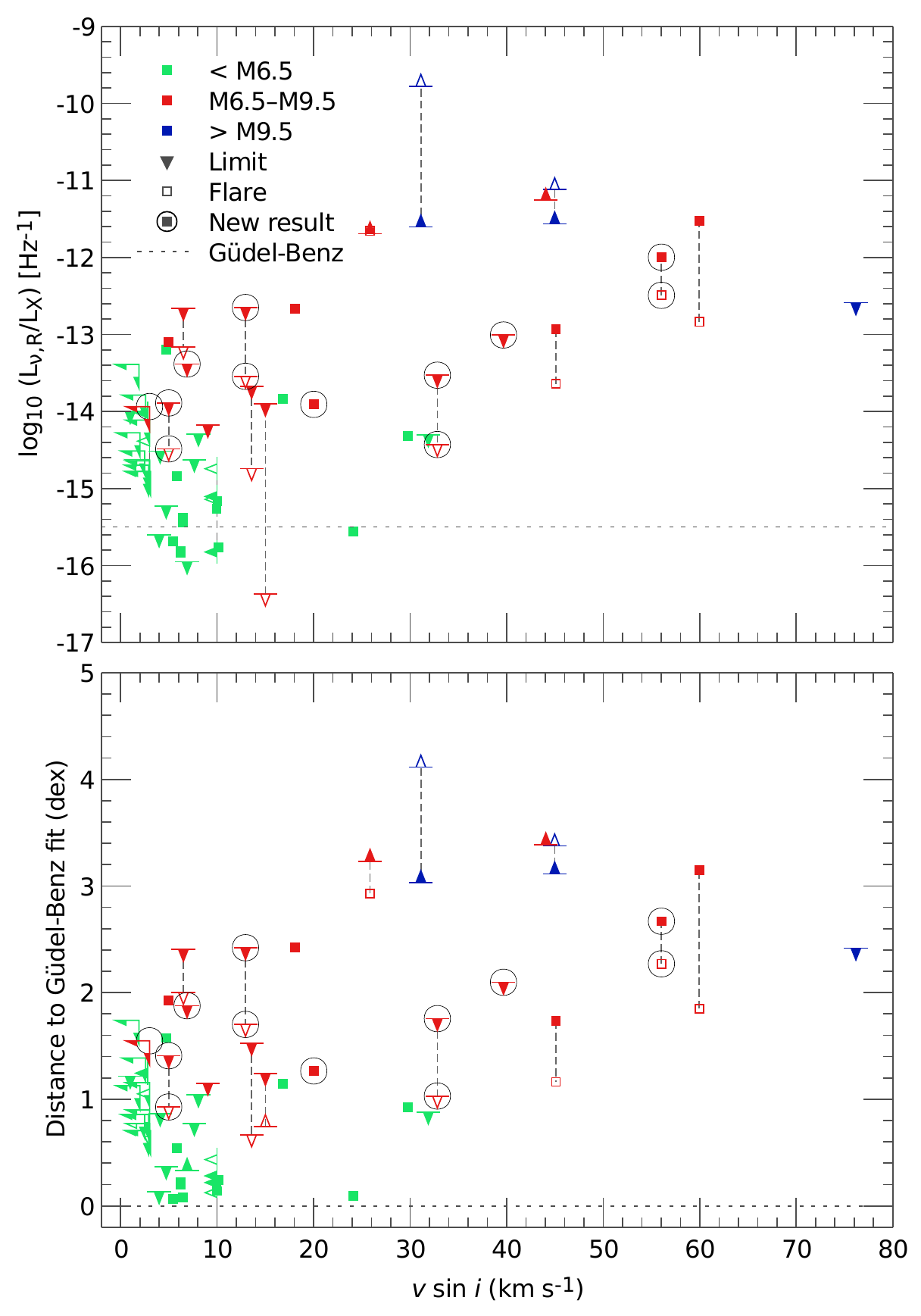}
  \caption{\textit{Upper panel}: measured ratios of radio to X-ray
    luminosities as a function of projected rotational velocity, \vsi. Colors
    and symbols are as in Figure~\ref{f.lxspt}. \textit{Lower panel:} same
    underlying data as the upper panel, but now plotting distance from the
    linear fit to the G\"udel-Benz relationship (Equation~\ref{e.gbfit}) rather
    than \sLrLx.}
  \label{f.rxrot}
\end{figure}

In Figure~\ref{f.rxspt} we plot \sLrLx\ as a function of spectral type.
\emph{There is clear evidence for new behavior at spectral types $\ge$M7}:
while there are no measurements of $\slrlx > -13$ in objects earlier than M7,
there are seven in later objects. The lower panel plots distance from the GBR
linear fit (Equation~\ref{e.gbfit}) rather than the simple ratio \sLrLx. As
can be seen, the choice of ordinate does not significantly affect the
structure of the data, and we discuss trends in terms of \sLrLx\ because that
quantity is closer to the observables.

In both panels of Figure~\ref{f.rxspt}, divergence from the GBR seems to
increase unconditionally with spectral type. This seeming trend is partially
misleading because of radio sensitivity limitations. While there is new
behavior at spectral types $\ge$M7, a population of $\ge$M7 objects consistent
with the early- and mid-M~dwarfs ($\slrlx\sim-15.5$) is not excluded. The new
observations presented in this work, which are of nearby objects and were
obtained with a highly sensitive radio telescope (the upgraded VLA), reach
limits of $\lr\sim12.5$. As shown in Figure~\ref{f.lxlr}, our data include the
most sensitive upper limit on M~dwarf radio emission available, $\slr < 11.8$
for \obj{lhs292}. Nonetheless, in our sample $\lx \sim 26$, so that we are
insensitive to $\slrlx \lesssim -13.5$. A source obeying the GBR with $\lx =
26$ and $d = 10$~pc would have a radio flux density of $S_\nu \sim 30$~nJy,
accessible only to the proposed Square Kilometer Array \citep[SKA;][]{theska}.
Until the arrival of an SKA-class telescope, the only way to probe UCDs in
this regime will be through observations of objects at distances of
$\lesssim$3~pc, such as the recently-discovered L/T binary \obj{luhman16}
\citep{l13,bsl13arxiv}.

Another issue affecting Figure~\ref{f.rxspt} is the correlation between mass
and rotational velocity in main-sequence late-type stars \citeeg{ib08}, which
obscures the true physical process underlying the observed trend. We plot
\sLrLx\ as a function of \vsi\ in Figure~\ref{f.rxrot}. Similar to
Figure~\ref{f.rxspt}, \sLrLx\ appears to increase with \vsi\ but the trend
requires care in interpretation. Along with the limitations in sensitivity to
small values of \sLrLx, some objects with low values of \vsi\ may be rapid
rotators seen at low inclinations. In this context, it is striking that out of
13 objects with $\vsi > 20$~\kms, seven have $\slrlx > -13.5$; that is, the
fraction of radio-over-luminous objects is large. The lack of
radio-over-luminous sources with small values of \vsi\ is consistent with the
argument of \citet{had+08} that viewing angle is an important factor affecting
the observed radio emission: if \sLrLx\ increases with $v$ but is independent
of $\sin i$, there should exist sources seen nearly pole-on with low values of
\vsi\ and large radio excess. Such sources have not yet been observed. The
presence of both lower and upper limits at high values of \vsi\ prevents a
simple characterization of the trend in \sLrLx. A more full analysis should
not only account for inclination effects, but also for source-to-source
variation. Studies of the relationship between X-ray emission and rotation
have shown that a useful parameter for doing so is the Rossby number,
$\text{Ro} = P_\text{rot} / \tau_\text{conv}$, where $P_\text{rot}$ is the
rotation period and $\tau_\text{conv}$ is the characteristic convective
overturn time \citep{nhb+84,pmm+03}. We pursue these matters in \papertwo.

\begin{figure*}[htbp]
  \plotone{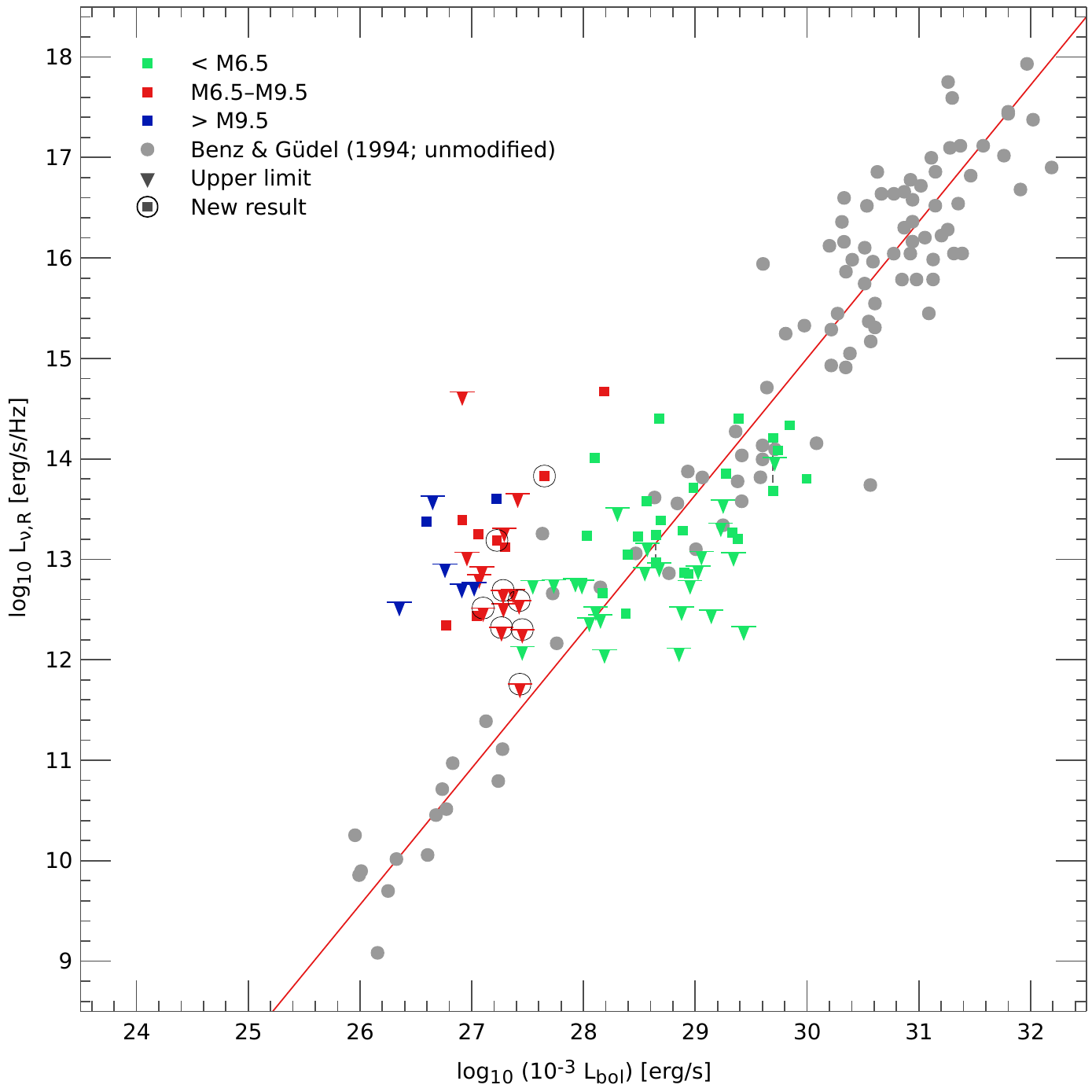}
  \caption{A ``pseudo G\"udel-Benz'' relationship between \sLr\ and the
    hypothetical ``saturated'' X-ray luminosity associated with each star,
    $\Lxsat = 10^{-3} \Lb$. Colors and symbols are as in Figure~\ref{f.lxspt}.
    The data from \citet{bg94} (\textit{gray circles}) have not been modified.
    Even if UCDs (\emph{blue and red points}) emitted X-rays at typical dMe
    ``saturation'' levels, they would still be radio-over-luminous compared to
    the GBR. If $<$M6.5 dwarfs emitted X-rays at these levels, on the other
    hand, they would cluster around the GBR.}
  \label{f.lxlradj}
\end{figure*}

Finally, although we have discussed departure from the GBR in terms of ``radio
over-luminosity,'' it could clearly be expressed in terms of ``X-ray
under-luminosity'' as well, a framing well-motivated by the dropoff in
\LxLb\ seen in UCDs. This prompts us to consider a modification of the GBR in
terms of the benchmark ``saturation'' level of X-ray emission, defining
$\Lxsat = 10^{-3} \Lb$. In Figure~\ref{f.lxlradj} we plot the same radio data
as in Figure~\ref{f.lxlr} but replace \Lx\ with \Lxsat. We omit flaring
measurements and do not alter the data from \citet{bg94}. As should be
expected from Figure~\ref{f.lxspt}, the UCD measurements move closer to the
canonical GBR. The data for objects $<$M6.5 straddle the canonical relation
nicely, with a substantial subset of measurements that are radio
\emph{under}luminous. The UCDs, however, remain displaced to the left of the
GBR; in other words, even if they emitted X-rays at the canonical
``saturation'' level, they would still be over-luminous in the radio. Any
explanation of GBR divergence in UCDs must therefore not merely account for
the suppression of X-ray emission relative to the stellar ``saturation''
trends, but also account for a comparative increase in radio emission. For
instance, if the suppression of UCD X-ray emission is entirely due to less
efficient heating of the corona leading to a temperature appreciably lower
than the typical coronal value of $\approx$1~keV \citeeg{bgg+08}, a mechanism
must still be proposed to explain the unexpectedly bright radio emission. We
argue below that in the radio-over-luminous sources, \sLr\ no longer scales
with \Lb.

\subsection{Summary of Observed Trends}

The X-ray emission of UCDs drops off rapidly with spectral type, with slightly
different properties depending on whether \Lx\ or \LxLb\ is considered
(Figure~\ref{f.lxspt}). Some UCDs diverge strongly from the GBR, with values
of \sLrLx\ exceeding the typical value of $10^{-15.5}$ by 4 orders of
magnitude; others may be consistent with it, with the limited sensitivity of
radio observations allowing us only to conclude that $\slrlx \lesssim -13.5$
(Figure~\ref{f.lxlr}). Although variability is an important consideration,
extreme GBR divergence is seen in simultaneous radio and X-ray observations,
confirming its reality (Figure~\ref{f.lxlrsim}). More extreme divergence from
the GBR seems to become possible at later spectral types
(Figure~\ref{f.rxspt}). Even if UCD X-ray activity (\LxLb) did not drop off
rapidly with spectral type, but rather remained at the standard ``saturation''
level of $\lxlb = -3$, UCDs would still be radio-over-luminous compared to the
GBR (Figure~\ref{f.lxlradj}).

The interpretation of these trends is complicated by the correlation between
mass and rotational velocity in main-sequence late-type stars \citeeg{ib08}.
Only objects with $\vsi \gtrsim 20$~\kms\ are seen to diverge strongly from
the GBR, and $\approx$50\% of such objects do so (Figure~\ref{f.rxrot}).
Although we discuss the role of rotation below, we defer detailed
investigation of the relationship between rotation and magnetic activity in
UCDs to \papertwo.

\section{Discussion}
\label{s.disc}

Our results underscore the wide range of magnetic phenomenology seen in the
UCD regime. For instance, our new observations of the M7.5~dwarf \obj{lp851}
yield a radio over-luminosity of $\lesssim$2.5 orders of magnitude. Meanwhile,
simultaneous observations of the nearby M8.5~dwarf \obj{1835+32} have revealed
a radio over-luminosity of $\gtrsim$5 orders of magnitude
\citep{bbg+08,had+08}; this assessment is relative to its mean radio emission
(\apx0.5~mJy), not the bright, highly polarized pulses that it has also been
seen to emit \citep[\apx2~mJy;][]{had+08}. Both objects are relatively rapid
rotators, with $\vsi = 33$ and $50$~\kms, respectively. (We emphasize that
while inclination effects may cause a rapid rotator to appear as a slow
rotator, they cannot cause a slow rotator to appear as a rapid rotator.) Both
are nearby \citep[9.7 and 5.7~pc;][]{cpbd+05,rcl+03}, and neither is known to
have a companion \citeeg{fbc+09,scc+05}. Despite these similarities, our new
data show that while \obj{lp851} is least an order of magnitude brighter than
\obj{1835+32} in the X-ray, it is also at least an order of magnitude fainter
in the radio.

\citet{sab+12} considered the GBR in the UCD regime and proposed the existence
of two populations: one comprising rapidly rotating, radio-bright, X-ray-dim
UCDs; the other comprising slower-rotating, radio-dim, X-ray-bright objects.
They additionally noted that only the radio-bright objects display bright,
highly polarized radio pulses, while only the radio-dim objects produce X-ray
flares, with the exception of \obj{lp944} which seems to flare in both bands.
Our data show that rotational velocity is not strictly tied to this dichotomy.
\obj{lp851} and \obj{n40026} are both rapid rotators (33 and 40~\kms,
respectively) that are nonetheless radio dim and X-ray bright. The late-M
binary \obj{n33370} is also a rapid rotator \citep[45~\kms;][]{mbi+11} that
has relatively low divergence from the GBR, being radio over-luminous by
\apx2.5 orders of magnitude in quiescence (Williams et al., in
preparation).

We propose that the range of observed UCD behavior may be due to varying
topology of their magnetic fields. In their study of the relationship between
rotation and radio activity in UCDs, \citet{mbr12} proposed such a connection;
here, we extend it to include X-ray emission as well. The connection to
magnetic topology is motivated by a series of studies of M~dwarfs using
Zeeman-Doppler imaging (ZDI) techniques \citep{dmp+08, mdp+08, mdp+10}. In
ZDI, the magnetic field topology is reconstructed from measurements of the
Zeeman effect in time-resolved optical spectropolarimetry
\citep{thezdi}\footnote{While the Zeeman effect provides a direct measurement
  of the field, it is important to note that the quantity being measured is
  the net \emph{signed} field in each resolution element. FeH spectroscopy,
  discussed below, probes the magnitude but not the topology of the
  \emph{unsigned} field.}. \citet{mdp+10} find that some M~dwarfs have strong,
axisymmetric fields, while others have fields that are weak and
non-axisymmetric. While early-M~dwarfs inhabit the weak-field regime and
mid-M~dwarfs inhabit the strong-field regime, late-M dwarfs inhabit
\emph{both} regimes. This observational finding led \citet{mdp+10} to propose
two magnetic dynamo modes leading to differing magnetic topologies, with
late-M dwarfs having a bistable dynamo that may inhabit either mode. This
concept is supported by recent results from geodynamo simulations that show
bimodal dynamo outcomes in rapid rotators \citep{mdsd11,gmd+13}.

We suggest that UCDs with strong, axisymmetric fields tend to have values of
\sLrLx\ in line with the GBR, while the ones with weak, non-axisymmetric fields
are radio over-luminous. Slow rotators ($\lesssim$20~\kms) have strong-field
dynamos and thus stay close the GBR, as found by \citet{sab+12}. Rapid
rotators may have dynamos in either mode, so some are strong GBR violators
while others are not, explaining the results of \citet{mbr12} and this work.
The association of sources that stay near the GBR with strong fields provides
continuity with the ZDI and X-ray observational results for mid-M~dwarfs. The
similar patterns seen in the ZDI topology and \sLrLx\ results are the primary
motivation for our hypothesis.

\begin{figure}[htbp]
  \plotone{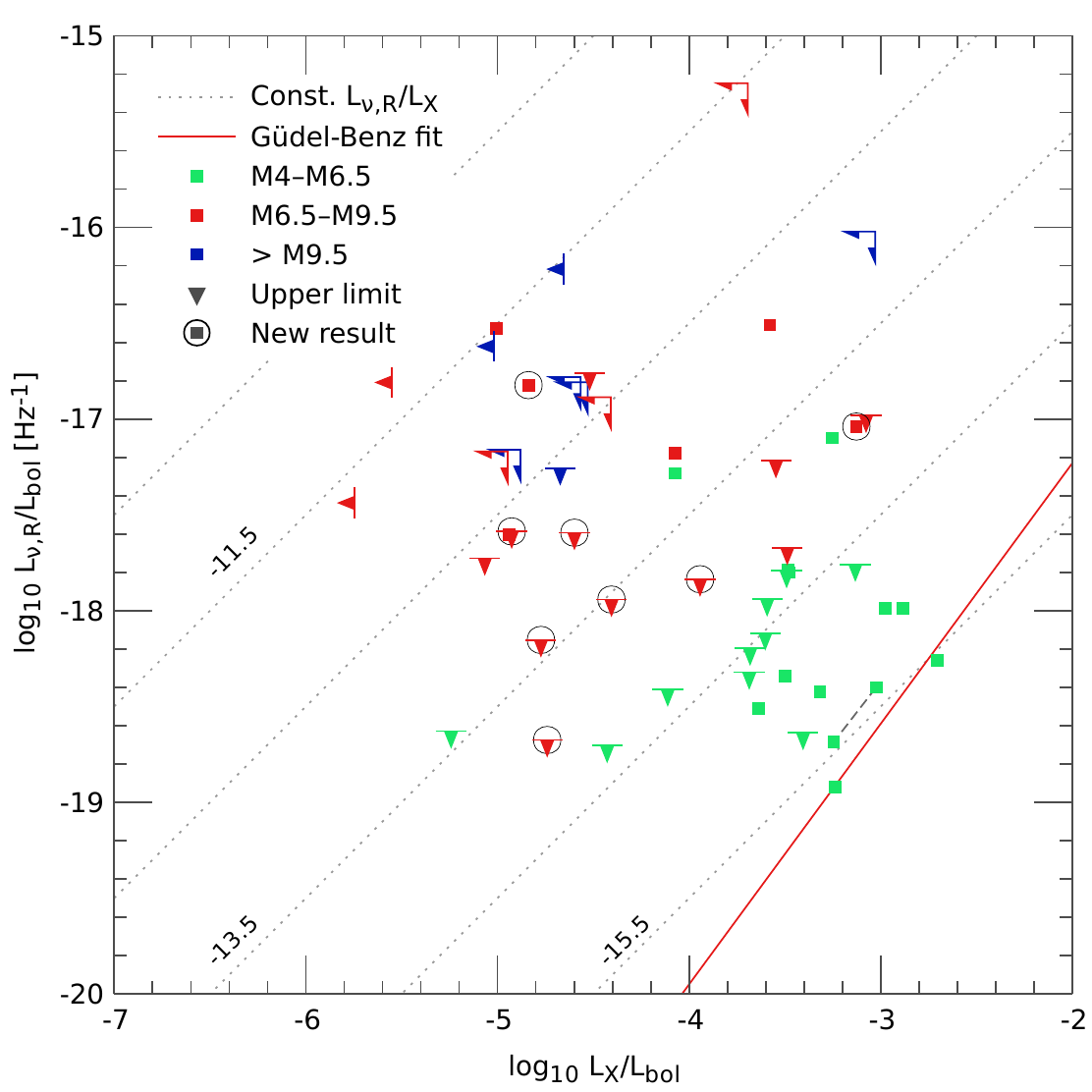}
  \caption{\sLrLb\ against \LxLb\ in fully-convective cool dwarfs. Symbols are
    as in Figure~\ref{f.lxlr}, except that \emph{green points} include only
    M4--M6.5 dwarfs. \emph{Dotted diagonal lines} indicate constant values of
    \sLrLx. \emph{Red line} matches the slope of the fit to the G\"udel-Benz
    relation in Figure~\ref{f.lxlr} and passes through the point representing
    Gl~166~C (M4.5), which has simultaneous radio and X-ray observations and
    lies extremely close to the fit given in Equation~\ref{e.gbfit}.}
  \label{f.rbxb}
\end{figure}

We have described the difference between the proposed strong-field and
weak-field sources in terms of \sLrLx, and argued that the weak-field sources
are radio over-luminous and X-ray under-luminous. However, the weak-field
sources also tend to have later spectral types, and as shown in
Figure~\ref{f.lxspt} these sources have lower values of \Lx. Some component of
the X-ray under-luminosity of the weak-field sources may be due to their
cooler temperates rather than the proposed effects of field topology. To
clarify this issue, we plot in Figure~\ref{f.rbxb} a modified G\"udel-Benz
relation in which we restrict the data to fully-convective dwarfs and
normalize by \Lb, thereby rearranging the data points along lines of constant
\sLrLx. It is noteworthy that even after accounting for changes in \Lb, the
radio-bright sources are still X-ray underluminous compared to the radio-dim
ones. Hypothetically, if the mean \Lb\ of the radio-bright objects were
\emph{much} lower than that of the radio-dim objects, the two groups might
have indistinguishable values of \LxLb, and the scatter in Figure~\ref{f.rbxb}
might be narrow in the \LxLb\ direction and broad in the \LrLb\ direction.
This is not the case. Instead, even after taking the changing \Lb\ into
account, the radio-bright sources seem to require both smaller \LxLb\ and
larger \sLrLb. We argue below that for the radio-bright sources \sLr\ is in
fact a more appropriate quantity to consider than \sLrLb, whereas for the
radio-faint ones the opposite is true.

The obvious issue to consider in the bimodal dynamo scenario is how the two
field topologies lead to such different values of \sLrLx. We start first with
the weak-field, radio-bright mode. It is plausible that this mode extracts
energy from stellar convective motions less efficiently than the strong-field
mode, leading to a decrease in overall magnetic activity that manifests itself
in decreased \LxLb\ compared to the strong-field objects. Meanwhile,
Figure~\ref{f.lxlr} shows that these objects tend to have similar values of
$\slr \approx 13.5$. We conjecture that in this mode, tangled magnetic field
structures lead to persistent small-scale reconnection events that are
sufficient to maintain a population of gyrosynchrotron-emitting electrons
filling a coronal region a few stellar radii ($R_*$) in size. This model has
already been suggested for the radio-bright sources \citeeg{bbg+08, brpb+09,
  mbi+11}. Because the radio luminosity is energetically unimportant compared
to \Lx\ or \Lb, the radio-emitting population could be maintained even at very
low levels of magnetic activity. The observed spectral luminosity
$\sLr\propto\nu^2 T_b R^2$, where $T_b$ is the brightness temperature and $R$
is the characteristic size of the emitting region. In the UCD regime, $R_*
\apx R_J$ with very weak dependence on mass. Supposing $R \apx R_*$, we find
that in the radio-bright population $T_b \apx 10^{8.5{-}9.5}$~K. These results
are consistent with values found in the dM1e binary YY~Gem, for which $T_b =
1.1 \times 10^9$~K and $R \apx 2 R_*$ were derived from VLBI observations
\citep{abg97}. Taking $T_b$ to be insensitive to mass, $\slr \approx 13.5$
should be a characteristic value for radio-bright UCDs observed at 8.5~GHz.
\citet{had+08} pointed out the relatively stable values of \sLr\ across the M
spectral type and argued for a different emission mechanism (see below). Our
interpretation differs in that we argue that emission from early-M~dwarfs is
instead due to magnetic reconnection in the standard chromospheric heating
picture, leading to consistency with the GBR. In this scenario, the similar
values of \sLr\ in the early-M~dwarfs and the radio-bright UCDs are
coincidental.

In the strong-field mode, we hypothesize that the coupling to internal stellar
convective motions is stronger, resulting in comparatively higher levels of
magnetic activity and \LxLb. These are nonetheless lower than what is found in
rapidly-rotating earlier-type stars, an effect often attributed to the outer
layers of UCD atmospheres becoming appreciably neutral, reducing their
coupling to the coronal magnetic field and thus their ability to inject energy
into it through surface convective motions \citep{mbs+02}. The standard
chromospheric evaporation model still applies, leading to values of
\sLrLx\ compatible with the GBR. However, these objects are apparently unable
to sustain a corona-filling population of gyrosynchrotron-emitting particles,
perhaps because reconnection events are simply too rare. The rarer
reconnection events would be more energetic than those seen in the weak-field
sources, perhaps explaining the observation of \citet{sab+12} that the
radio-quiet sources are seen to flare in the X-ray while the radio-loud ones
are not. This may also explain the conjecture of \citet{rps10} that some UCDs
exhibit large X-ray flares and low-level X-ray variability but, unlike active
early-M~dwarfs, not a continuous spectrum of flare energies.

\begin{figure}[htbp]
  \plotone{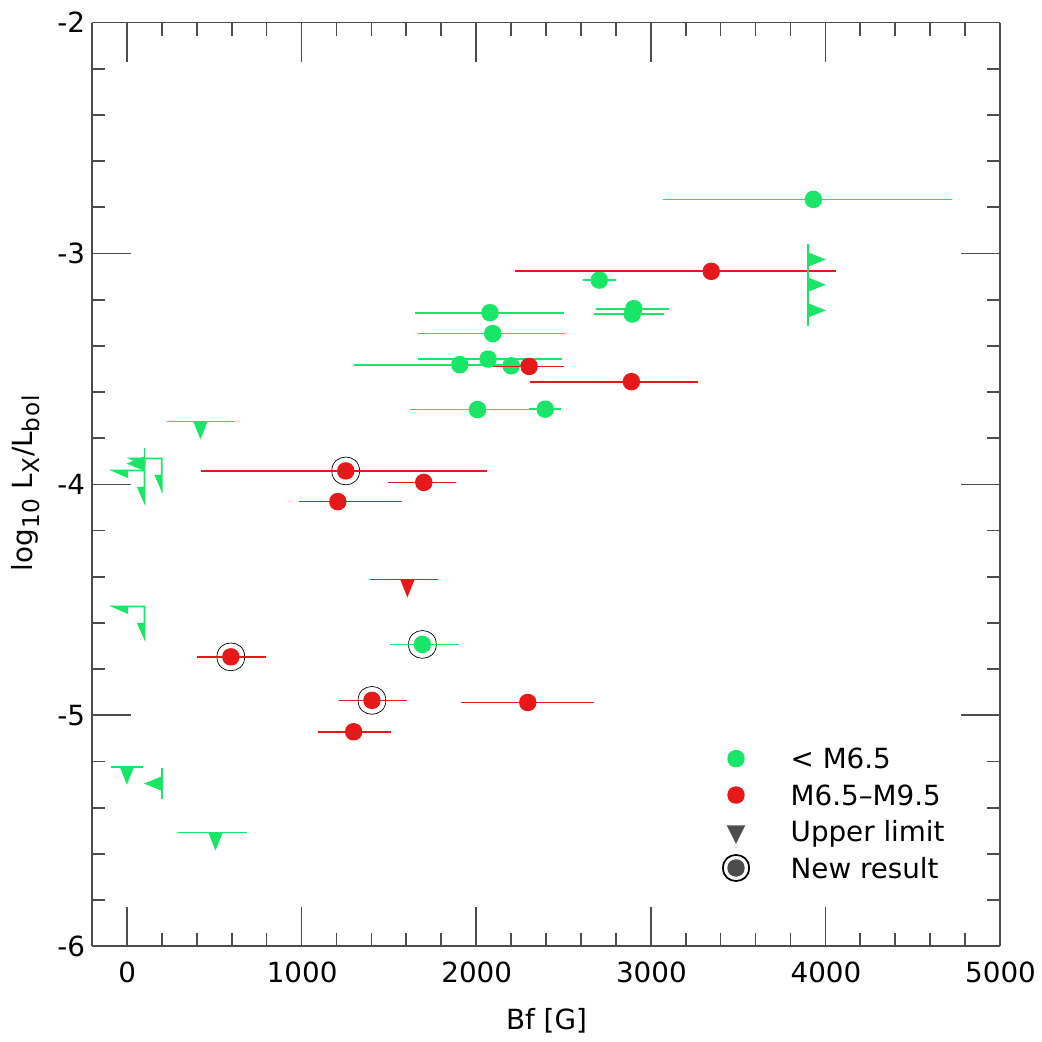}
  \caption{X-ray activity as a function of FeH-diagnosed disk-averaged field
    strength $Bf$ for UCDs (\emph{red}) and earlier M~dwarfs (\emph{green}).
    The $Bf$ data are from \citet{rb07,rb10} and \citet{rbb09}, while the
    \LxLb\ data are from Table~\ref{t.dbdata} excluding known flares. With the
    exception of \obj{1048-39} \citep[M9; $\lxlb\apx-5$;][]{sab+12}, all
    objects with $Bf>2$~kG ($< 2$~kG) have $\lxlb > -4$ ($< -3.5$).}
  \label{f.bflx}
\end{figure}

This model posits that UCDs with stronger magnetic fields should be associated
with higher values of \LxLb. Figure~\ref{f.bflx} shows that this is
qualitatively the case when the field strength is diagnosed with FeH
spectroscopy \citep{rb06}, a technique that has the advantage of having been
applied to numerous cool dwarfs \citep{rb07,rbb09,rb10}. This technique
measures $Bf$, the average unsigned magnetic field at the photosphere with
$f\le1$ being a filling factor term. ZDI, being sensitive to the net signed
field in each resolution element, measures strictly smaller values. Despite
this distinction, the two measures are generally correlated, with ZDI values
being \apx5--15\% of FeH values depending on mass \citep{rb09b,mdp+10}. A
correlation similar to that in Figure~\ref{f.bflx} exists between $Bf$ and
$\LhLb$ \citep{rb10}. Figure~\ref{f.rbxb} shows that the very radio-bright
UCDs have $\lxlb \lesssim -4.5$; the five UCDs in Figure~\ref{f.bflx} meeting
this criterion (\obj{1048-39}, \obj{lhs292}, \obj{lhs2924}, \obj{lp647},
\obj{vb10}) have relatively low values of $Bf \lesssim 2.5$~kG. Three UCDs
have $\lxlb \approx -4$ and $Bf < 2.5$~kG (\obj{lhs248}, \obj{lhs3003},
\obj{lhs3406}), consistent with the idea that both field topology and field
strength are important parameters. All of the objects with $Bf > 2.5$~kG have
$\lxlb > -4$.

A striking characteristic of the radio-bright UCDs is that several of them
emit periodic, bright, highly polarized radio pulses \citep{had+06, had+08,
  bgg+08, brpb+09}. These are characteristic of the electron cyclotron maser
instability \citep[ECMI;][]{theecm,t06}, a coherent process that results in
radio emission at the cyclotron frequency $\omega_c = e B / m c$. This process
is responsible for auroral radio bursts in the magnetic solar system planets,
which have many similarities to those observed in the UCDs \citep{z98}. In
particular, the periodic nature of the pulses is strongly suggestive of beamed
emission tied to the stellar rotation. We suggest that, despite their
generally tangled fields, the weak-field sources are able to maintain the
large-scale converging magnetic structures needed to produce these auroral
bursts. These structures need not dominate the ZDI-diagnosed field since their
footpoints may cover only a small fraction of the photosphere; hence
their possible existence is not inconsistent with the overall finding of a
weak, non-axisymmetric field. (In \citet{mdp+10}, the ratio of
the maximum and mean ZDI-diagnosed fields, $B_\text{max} / \langle B\rangle$,
can exceed 5, and $B_\text{max}$ may be underestimated if the filling factor
of the relevant field structure is low.)
We have
argued that the weak-field sources maintain a population of energetic
electrons filling the coronal volume, providing a continual supply of
particles capable of driving the bursts as well. The strong-field sources are
likely also able to maintain the requisite converging fields but, lacking the
corona-filling supply of particles, do not display the periodic bursts. As
pointed out by \citet{sab+12}, however, the strong-field sources are slower
rotators, and so the lack of observed bursts may be a selection effect because
UCD radio observations are generally shorter in duration than the rotational
period. \citet{had+08} identify another possible selection effect, providing
evidence that the sources with ECMI bursts are observed at high inclinations.

\citet{had+06,had+08} further argue that the GBR violators are radio bright
because the vast majority of their radio emission is due to the ECMI,
including the non-burst emission that has generally been attributed to
gyrosynchrotron processes \citeeg{b02,b06b}. They suggest that depolarization
and steady particle acceleration could cause ECMI emission to have the low
variability and polarization that are associated with the non-burst emission
of these sources. While this argument may apply to some UCDs, we disfavor its
application to \emph{all} of them, because it calls for an implausible
physical configuration in sources such as \obj{0746+20}. This L~dwarf binary
periodically emits rapid (\apx100~s) pulses of \apx10~mJy and \apx100\%
circular polarization atop seemingly quiescent emission at \apx0.3~mJy with
$\lesssim$15\% circular polarization and a spectral index of $\alpha \apx
-0.7$ ($S_\nu \propto \nu^\alpha$) \citep{brpb+09}. Any variation besides the
pulses is at the $\lesssim$1~mJy level. If the quiescent emission also
originates in the ECMI, two emitting regions of identical field strengths
would be required, both with the physical conditions suitable for ECMI
cascade, but producing vastly different observed emission. It seems simpler to
posit that both ECMI and gyrosynchrotron emission occur in this case. In other
sources such as \obj{0036+18}, both polarized and unpolarized emission of
comparable flux densities and variability characteristics are observed, and
the arguments of \citet{had+06,had+08} are more persuasive.

In our scenario it is still not entirely clear what factors determine the
overall level of magnetic activity, i.e. \LxLb. There are substantial
correlations among \LxLb, spectral type (Figure~\ref{f.lxspt}), GBR deviation
(Figure~\ref{f.rbxb}), and rotational velocity (\papertwo). In the quest to
understand these interconnections, two particular issues call for
investigation. The first is to what extent the observed UCD
``supersaturation''-like effect, an anticorrelation between \LxLb\ and
\vsi\ \citep{bbf+10}, is a causal relationship, relating perhaps to evolution
in the nature of the dynamo. The second is the question of which processes
drive the observed dropoff in \LxLb\ as a function of spectral type
(Figure~\ref{f.lxspt}), relating perhaps to rotation, changes in internal
structure, or increased photospheric neutrality \citep{mbs+02}. Because of the
correlations among the variables in question, the role of rotation must be
considered carefully in attempts to resolve these issues, and so we defer
further analysis to \papertwo.

Finally, we wish to emphasize that although we have discussed a model of two
distinct dynamo modes, these modes are not necessarily mutually exclusive
within a given source -- it is plausible that one object could host multiple
field-generating processes. With a sample of 29 sources and many
non-detections, it would be premature to claim that there are two distinct
subpopulations.

\section{Summary and Conclusions}
\label{s.conc}

We presented new X-ray (\chandra) and radio (VLA) observations of seven
ultracool dwarfs with spectral types between M6.5 and M9.5 and a wide range of
\vsi. We have detected all of them in the X-ray band, nearly doubling the
number of UCDs with X-ray detections. Despite the increased sensitivity of the
upgraded VLA, only one of the sources was detected in the radio. Our results
are thus broadly consistent with the G\"udel-Benz relationship between radio
and X-ray emission in stellar flare phenomena, though they still admit radio
over-luminosities of a factor of \apx$10^2$. This is in contrast to several
spectacular recent results, in which some UCDs are seen to be
radio-over-luminous by \apx5 orders of magnitude. Although UCDs are highly
variable in both bands, making simultaneous observations important tools in
the analysis of correlations such as the GBR, we have argued that the
nonsimultaneity of our observations does not significantly affect their
interpretation.

We have also assembled a comprehensive sample of UCDs with both radio and
X-ray observations (Figure~\ref{f.lxlr}; Table~\ref{t.dbdata}), including all
known measurements in which the observations were simultaneous. With the
addition of our new measurements, there is strong evidence that UCDs display a
wide range of behavior with regards to the GBR: some are strongly radio
over-luminous, while others could be consistent with it. This range can be, and
has been, interpreted as a bimodality in the UCD population
\citep{mbr12,sab+12}, which has support from both ZDI observations and
geodynamo simulations \citep{mdp+10,gmd+13}. We have argued that one group of
sources can maintain a population of gyrosynchrotron-emitting particles in the
corona, setting an effective floor on \sLr, while the other group is less
active than earlier-type stars but emits following the standard chromospheric
evaporation model. Interpretation of the data, however, is made difficult by
the many variables at play: $M_*$, \teff, \vsi, age, metallicity, binarity,
and long- and short-term variability. It is not clear that the population can
be neatly divided into two distinct subgroups at this time.

Although the study of UCD magnetism continues to present many puzzles, we see
several reasons to be optimistic for the future. Studies of progressively
larger samples will help clarify trends and allow more robust examination of
subsamples that control for variables such as mass, age, or rotation. The
upgrade of the VLA presents a major opportunity in this regard, because it is
becoming clear that radio observations present the best opportunity for
exploring the magnetism of the coolest objects, whose faintness and rapid
rotation make extremely difficult the application of techniques such as ZDI.
Furthermore, new benchmark objects are being discovered that offer the chance
for detailed study. These include the radio-brightest UCD, \obj{n33370}
\citep[Williams et al., in preparation]{mbi+11}, and the coolest UCD yet
detected in the radio, \obj{1047+21} \citep[T6.5;][]{rw12,wbz13}. Wide-field
infrared surveys and proper motion searches are discovering substantial
numbers of UCDs, including extremely nearby examples such as the 2-pc L/T
binary \obj{luhman16} \citep{l13,bsl13arxiv}. Finally, theoretical studies of
the operation of dynamo action in fully convective bodies are progressing,
with numerical geodynamo models being adapted to be able to simulate density
contrasts closer to those seen in UCDs \citeeg{chr09}. Advances in all of
these areas will greatly increase our understanding of the magnetism of brown
dwarfs and, eventually, extrasolar planets.

\acknowledgments

We thank Scott Wolk and Katja Poppenhaeger for enlightening discussions. We
also thank the anonymous referee for helpful, constructive comments. This
work is supported in part by the National Science Foundation REU and
Department of Defense ASSURE programs under NSF Grant no. 1262851 and by the
Smithsonian Institution. E.~B. and P.~K.~G.~W. acknowledge support for this
work from the National Science Foundation through Grant AST-1008361, and from
the National Aeronautics and Space Administration through Chandra Award Number
GO2-13007A issued by the Chandra X-ray Observatory Center, which is operated
by the Smithsonian Astrophysical Observatory for and on behalf of the National
Aeronautics Space Administration under contract NAS8-03060. The VLA is
operated by the National Radio Astronomy Observatory, a facility of the
National Science Foundation operated under cooperative agreement by Associated
Universities, Inc. This research has made use of the SIMBAD database, operated
at CDS, Strasbourg, France, and NASA's Astrophysics Data System.

Facilities: \facility{\chandra}, \facility{Karl G. Jansky VLA}

\pwiflocal%
  \bibliographystyle{../yahapj/yahapj}%
  \bibliography{../bib/pkgw,extra}{}%
\else%
  \bibliographystyle{yahapj}%
  \bibliography{pkgw,extra}{}%
\fi

\appendix

Here we give our method for computing bolometric luminosities and discuss
the properties of the sources for which we present new results.

\subsection{Bolometric luminosities}

We compute bolometric luminosities using bolometric corrections to observed
absolute magnitudes. In particular,
\begin{equation}
\lb = \frac{2}{5} (M_{\odot,\text{bol}} - M - \text{BC}) + [L_\odot],
\end{equation}
where $M$ is the absolute magnitude of the star in some band, BC is the
bolometric correction in that band, and $M_{\odot,\text{bol}} = 4.7554$ is the
bolometric absolute magnitude of the Sun. For M dwarfs, we use the average of
the bolometric luminosities calculated using $J$ and $K$ band magnitudes:
\begin{equation}
  \text{BC}_K = 2.43 + 0.0895\times (\text{SP})
\end{equation}
and
\begin{equation}
  \text{BC}_J = 1.53 + 0.148\times(\text{SP}) - 0.0105\times(\text{SP})^2,
\end{equation}
with SP = 0 (5) for spectral type M0 (M5) \citep{wgm99}. For L dwarfs, we use
the $K$-band magnitude only:
\begin{equation}
\text{BC}_K = \begin{cases}
  2.37 + 0.075\times(\text{SP}), & 10 \leq \text{SP} < 14, \\
  4.47 - 0.075\times(\text{SP}), & 14 \leq \text{SP} \leq 19,
\end{cases}
\end{equation}
with SP used consistently as above (SP = 10 for L0) \citep{nty04}.

\subsection{Properties of Newly-Observed Targets}

\subsubsection{\objn{lhs292}}

\obj{lhs292} (= GJ\,3622, LP\,731--58) is a nearby ($d \approx
4.5$~pc) M6.5~dwarf. It is a variable \ha\ emitter with $\lhlb \approx
-4.4$ \citep{dlh86,lbk10,rb10} and has been suggested as a candidate UV~Cet
flare star \citep{dlh86}. $K$-band spectroscopy and modeling yield an
estimated $\teff = 2772\pm25$~K and $[\text{Fe}/\text{H}] = -0.41\pm0.17$
\citep{racml12}. \citet{rb10} used other spectroscopic measurements to
determine $\vsi < 3\pm2$~\kms\ and $Bf = 600\pm200$~G.
ROSAT observations by \citet{fgsb93} yielded an upper limit of $\lx < 26.42$
in the 0.1--2.4~keV band. \citet{mbr12} observed \obj{lhs292} in the radio and
did not detect it, obtaining a limit of $S_\nu < 96$~\ujy\ ($\lrlb <
-8.20$) at 8.46~GHz. \citet{mdp+10} used Zeeman-Doppler imaging to derive a
low-resolution map of the stellar magnetic field structure. Although the S/N
was low, their modeling indicates a dipolar field of \apx100~G with an
inclination $i \approx 60\degr$. From variability in its radial velocity on
the 10~\kms\ level, \citet{gw03} argue that \obj{lhs292} is a binary system,
but a companion has not yet been detected by other means.

\subsubsection{\objn{lhs523}}

\obj{lhs523} (= GJ\,4281, LP\,760--3) is an M6.5~dwarf at a
distance of 11~pc. \citet{gl86} measured its \ha\ EW to be 1.3~\AA.
\citet{mb03} measure an \ha\ EW of 4.4~\AA\ ($\lhlb = -4.15$) and $\vsi =
7.0$~\kms, consistent with a later determination of $\vsi < 12$~\kms\ using a
different method \citep{ljp+12}. A proper motion of $1200\pm130$ mas~yr$^{-1}$
has been determined \citep{dhc05}. \citet{jrj+09} determine $\teff = 2536$~K
and $M = 0.093\pm0.005$~M$_\sun$ from photometric modeling. \obj{lhs523} was
undetected in a ROSAT survey of very-low-mass stars \citep{fgsb93}, with $\lx
< 26.89$.

\subsubsection{\objn{lhs2397}}

\obj{lhs2397} (= GJ\,3655, LP\,732--94) is an M8$+$L7.5 binary at a distance
of 14~pc. The brown dwarf companion was detected by direct imaging with
adaptive optics by \citet{fcs03}. Long-term monitoring has led to a detailed
model of the binary orbit \citep{dli09,kgb+10}, with the total system mass
being estimated as $0.144 \pm 0.013$~\msun\ \citep{kgb+10}. The projected
rotational velocities of the primary and secondary have been measured as $\vsi
= 15\pm1$ and $11\pm3$, respectively \citep{kgf+12}. \citet{mb03} detected
strong \ha\ emission with EW = 29.4~\AA, computing $\lhlb = -3.70$. Comparable
results were obtained by \citet{scb+07}, who found even stronger emission with
$\lhlb = -3.40$, making it one of the most \ha-luminous UCDs known.

\subsubsection{\objn{lhs3406}}

\obj{lhs3406} (= V492~Lyr, GJ\,4073, LP\,229--30) is an M8~dwarf at a distance
of 14~pc. In an investigation of candidate cataclysmic variable systems,
\citet{lhlc99} confirmed its dMe nature and measured $\lhlb \approx -4.1$.
Subsequent activity observations by \citet{scb+07} find $\lhlb \approx -4.3$.
High-resolution spectra obtained by \citet{rb10} reveal a rotational velocity
of $\vsi = 5.0\pm3.2$~\kms, $\lhlb = -4.1$, and $Bf = 1200\pm800$~G. This is
consistent with the results of \citet{dmm+12}, who find $\vsi < 12$~\kms.
Previous radio observations with the VLA yielded an upper limit of $S_\nu <
48$~\ujy\ at 8.46~GHz \citep{b06b}. \obj{lhs3406} was undetected in the ROSAT
survey of \citet{fgsb93} with an upper limit of $\lx < 26.83$. In a search for
brown dwarf companions, no objects were found within limits of $\Delta J <
9.6$, $\Delta \theta > 10''$ \citep{mz04}.

\subsubsection{\objn{lp647}}

\obj{lp647} (= NLTT\,3868) is an M9~dwarf at a distance of 11~pc. In the
survey of \citet{scb+07}, it was found that $\lhlb \approx -4.5$. \citet{rb10}
found a rotational velocity of $\vsi = 13\pm2$~\kms\ as well as $\lhlb =
-4.50$ and $Bf = 1400\pm200$~G. \citet{dmm+12} obtain consistent results, with
$\vsi < 12$~\kms. Radio observations with the VLA have yielded an upper limit
of $S_\nu < 33$~\ujy\ at 8.46~GHz \citep{b06b}.

\subsubsection{\objn{lp851}}

\obj{lp851} (= DENIS-P\,J115542.9$-$222458) is an M7.5~dwarf at a distance of
9.7~pc. Despite the fact that it has been suspected to be a very nearby object
for nearly 40 years \citep{thelhs}, it is poorly-studied. \citet{rb10} found a
rotational velocity of $\vsi = 33\pm3$~\kms\ and $\lhlb = -4.58$. Radio
observations with the VLA have yielded an upper limit of $S_\nu < 90$~\ujy\ at
8.46~GHz \citep{mbr12}.

\subsubsection{\objn{n40026}}

\obj{n40026} (= LSPM\,J1521$+$5053) is an M7.5~dwarf at a distance of 16.1~pc.
\citet{scb+07} measured $\lhlb \approx -4.9$. \citet{rb10} measured $\vsi =
40\pm4$~\kms\ and $\lhlb = -4.88$. Radio observations with the VLA have
yielded an upper limit of $S_\nu < 39$~\ujy\ at 8.46~GHz \citep{b06b}.
\citet{scc+05} found no evidence for any companions to \obj{n40026} between
0.1--15~arcsec with a separation-dependent contrast of $\Delta H \lesssim
12$~mag.

\subsection{Properties of Targets with New Analysis}

\subsubsection{\objn{g208-44-45}}

The \obj{g208-44-45} system (= GJ\,1245\,ABC) is a well-studied, nearby
($4.6$~pc), cool triple \citep{mhf+88}. \obj{g208-44} (= GJ\,1245\,AC,
LHS\,3494\,AB, LSPM\,J1953$+$4424W) is separated from \obj{g208-45} (=
GJ\,1245\,B, LHS\,3495, LSPM\,J1953$+$4424E) by \apx7$''$, and is itself a
tighter binary of \apx1$''$ separation. Both components of the overall system
are flare stars \citep{rcv80} and it has long been a target of activity
studies \citeeg{wjk89}. The blended pair \obj{g208-44} has a radio flux
density $<$192~\ujy\ at 5~GHz \citep{bbfk09}, $\lxlb \approx -3.78$
\citep{sfg95}, $\lhlb \approx -4.3$ \citep{mb03}, and $\vsi = 17.4 \pm
1.4$~\kms\ \citep{dfpm98}. Resolved measurements of \obj{g208-44} are
uncommon, with \citet{lhm08} recently providing the first spectral type
estimate for \objt{g208-44}\,B (M8.5); it is the only component of the system
that fits our definition of being an ultracool dwarf. \citet{mb03} were able
to measure \vsi\ in \objt{g208-44}\,A ($22.5 \pm 2$~\kms) and \obj{g208-45}
($6.8 \pm 1.9$~\kms), but not \objt{g208-44}\,B.

\subsubsection{\objn{dp0255}}

\objf{dp0255} (= 2MUCD\,10158; hereafter \obj{dp0255}), an L8~dwarf at a
distance of 5~pc, was identified as an extremely cool object by
\citet{mdb+99}, with $\teff \approx 1500$~K \citep{rb08}. Its \ha\ emission is
extremely faint, $\lhlb < -8.28$ \citep{rb08}. Multiple measurements of its
\vsi\ yield either \apx40 or \apx60~\kms, depending on the analysis method
used, likely due to its unusually late type \citep{mb03,zomb+06,rb08}. Brown
dwarf companions with separations of \apx7--165$''$ are unlikely
\citep{cmp+11}. Optical monitoring suggests short- and long-term aperiodic
variations \citep{k13}.

\subsubsection{\objn{lp349}}

\obj{lp349} (= LSPM\,J0027$+$2219, NLTT\,1470) is an M8+M9 binary at a
distance of \apx10~pc. \citet{gmr+00} found moderate \ha\ activity in the
blended system, with $\lhlb = -4.52$. \citet{fbd+05} used adaptive-optics
imaging to reveal the binarity of the system, finding a separation of
\apx0.1$''$ with $\Delta K' = 0.26 \pm 0.05$~mag. A measurement of the
trigonometric parallax, $\pi = 75.82 \pm 1.62$~mas, has only recently become
available \citep{gc09}. \citet{rb10} found $\lhlb = -4.53$, in very good
agreement with \citet{gmr+00}, but were unable to assess the magnetic field
strength. \citet{pbol+07} discovered radio emission from the system with a
flux density at 8.5~GHz of $365 \pm 16$~\ujy. Both of the components are rapid
rotators, although their projected rotational velocities differ by \apx15\%:
$\vsi = 55 \pm 2$~\kms\ and $83 \pm 3$~\kms\ for the A and B components,
respectively \citep{kgf+12}. The total mass of the system is
$0.120^{+0.008}_{-0.007}$~\msun\ and it is also likely to be young, at
$140\pm30$~Myr \citep{dlb+10}.

\clearpage
\begin{turnpage}
\begin{deluxetable}{r@{\,}llr@{}lc@{\,}cr@{\,}r@{}lr@{\,}r@{}lr@{\,}r@{}lr@{\,}r@{}lr@{\,}r@{}lcc}
\tabletypesize{\footnotesize}
\tablecolumns{24}
\tablewidth{0em}
\tablecaption{Radio and X-ray Data for UCDs\label{t.dbdata}}
\tablehead{
\colhead{} & \colhead{2MASS Identifier} & \colhead{SpT} & \multicolumn{2}{c}{[$\Lb$]} & \colhead{St.} & \colhead{X-Ray Band} & \multicolumn{3}{c}{[$L_X$]} & \multicolumn{3}{c}{[$L_{\nu,R}$]} & \multicolumn{3}{c}{[$L_X/\Lb$]} & \multicolumn{3}{c}{[$L_{\nu,R}/\Lb$]} & \multicolumn{3}{c}{[$L_{\nu,R}/L_X$]} & \multicolumn{2}{c}{References}  \\
 \cline{23-24}  &  &  & \multicolumn{2}{c}{[$L_\sun$]} &  & \colhead{(keV)} & \multicolumn{3}{c}{[erg s$^{-1}$]} & \multicolumn{3}{c}{[erg s$^{-1}$ Hz$^{-1}$]} &  &  &  & \multicolumn{3}{c}{[Hz$^{-1}$]} & \multicolumn{3}{c}{[Hz$^{-1}$]} &  &  \\ \\
\multicolumn{1}{c}{} & \multicolumn{1}{c}{(1)} & \multicolumn{1}{c}{(2)} & \multicolumn{2}{c}{(3)} & \multicolumn{1}{c}{(4)} & \multicolumn{1}{c}{(5)} & \multicolumn{3}{c}{(6)} & \multicolumn{3}{c}{(7)} & \multicolumn{3}{c}{(8)} & \multicolumn{3}{c}{(9)} & \multicolumn{3}{c}{(10)} & \multicolumn{1}{c}{(X)} & \multicolumn{1}{c}{(R)}
}
\startdata
 & \object{10481258$-$1120082} & M6.5 & $-3$ & $.15$ & -- & $0.2$--$2.0$ &  & $25$ & $.7$ & ~~~~~$<$ & $11$ & $.8$ &  & $-4$ & $.7$ & $<$ & $-18$ & $.7$ & $<$ & $-13$ & $.9$ & $\star$ & $\star$ \\
 & \object{22285440$-$1325178} & M6.5 & $-3$ & $.13$ & -- & $0.2$--$2.0$ &  & $25$ & $.7$ & $<$ & $12$ & $.3$ &  & $-4$ & $.8$ & $<$ & $-18$ & $.2$ & $<$ & $-13$ & $.4$ & $\star$ & $\star$ \\
\P & \object{13142039$+$1320011 AB} & M7 & $-2$ & $.40$ & Q & $0.2$--$2.0$ &  & $27$ & $.6$ &  & $14$ & $.7$ &  & $-3$ & $.6$ &  & $-16$ & $.5$ &  & $-12$ & $.9$ & 5 & 5 \\
\P &  &  &  &  & F & $0.2$--$2.0$ &  & $28$ & $.3$ &  & $14$ & $.7$ &  & $-2$ & $.9$ &  & $-16$ & $.5$ &  & $-13$ & $.6$ & 5 & 5 \\
 & \object{14563831$-$2809473} & M7 & $-3$ & $.29$ & -- & $0.1$--$2.4$ &  & $26$ & $.2$ &  & $13$ & $.1$ &  & $-4$ & $.1$ &  & $-17$ & $.2$ &  & $-13$ & $.1$ & 7 & 8 \\
 & \object{16553529$-$0823401} & M7 & $-3$ & $.21$ & -- & $0.1$--$2.4$ &  & $26$ & $.9$ & $<$ & $12$ & $.7$ &  & $-3$ & $.5$ & $<$ & $-17$ & $.7$ & $<$ & $-14$ & $.2$ & 10 & 10 \\
 & \object{11554286$-$2224586} & M7.5 & $-3$ & $.32$ & Q & $0.2$--$2.0$ &  & $25$ & $.8$ & $<$ & $12$ & $.3$ &  & $-4$ & $.4$ & $<$ & $-18$ & $.0$ & $<$ & $-13$ & $.5$ & $\star$ & $\star$ \\
 &  &  &  &  & F & $0.2$--$2.0$ &  & $26$ & $.8$ & $<$ & $12$ & $.3$ &  & $-3$ & $.5$ & $<$ & $-17$ & $.9$ & $<$ & $-14$ & $.4$ & $\star$ & $\star$ \\
 & \object{15210103$+$5053230} & M7.5 & $-3$ & $.30$ & -- & $0.2$--$2.0$ &  & $25$ & $.7$ & $<$ & $12$ & $.7$ &  & $-4$ & $.6$ & $<$ & $-17$ & $.6$ & $<$ & $-13$ & $.0$ & $\star$ & $\star$ \\
 & \object{00275592$+$2219328 AB} & M8 & $-2$ & $.93$ & Q & $0.2$--$2.0$ &  & $25$ & $.8$ &  & $13$ & $.8$ &  & $-4$ & $.8$ &  & $-16$ & $.8$ &  & $-12$ & $.0$ & $\star$ & 8 \\
 &  &  &  &  & F & $0.2$--$2.0$ &  & $26$ & $.3$ &  & $13$ & $.8$ &  & $-4$ & $.3$ &  & $-16$ & $.8$ &  & $-12$ & $.5$ & $\star$ & 8 \\
 & \object{03205965$+$1854233} & M8 & $-3$ & $.29$ & Q & $0.3$--$8.0$ &  & $27$ & $.2$ & $<$ & $13$ & $.3$ &  & $-3$ & $.1$ & $<$ & $-17$ & $.0$ & $<$ & $-13$ & $.9$ & 14 & 8 \\
 &  &  &  &  & F & $0.3$--$8.0$ &  & $29$ & $.7$ & $<$ & $13$ & $.3$ &  & $-0$ & $.6$ & $<$ & $-17$ & $.0$ & $<$ & $-16$ & $.4$ & 14 & 8 \\
 & \object{11214924$-$1313084 AB} & M8 & $-3$ & $.36$ & -- & $0.2$--$2.0$ &  & $27$ & $.1$ &  & $13$ & $.2$ &  & $-3$ & $.1$ &  & $-17$ & $.0$ &  & $-13$ & $.9$ & $\star$ & $\star$ \\
 & \object{18432213$+$4040209} & M8 & $-3$ & $.16$ & Q & $0.2$--$2.0$ &  & $26$ & $.5$ & $<$ & $12$ & $.6$ &  & $-3$ & $.9$ & $<$ & $-17$ & $.8$ & $<$ & $-13$ & $.9$ & $\star$ & $\star$ \\
 &  &  &  &  & F & $0.2$--$2.0$ &  & $27$ & $.1$ & $<$ & $12$ & $.6$ &  & $-3$ & $.4$ & $<$ & $-17$ & $.8$ & $<$ & $-14$ & $.5$ & $\star$ & $\star$ \\
 & \object{19165762$+$0509021} & M8 & $-3$ & $.30$ & Q & $0.2$--$2.0$ &  & $25$ & $.2$ & $<$ & $12$ & $.6$ &  & $-5$ & $.1$ & $<$ & $-17$ & $.7$ & $<$ & $-12$ & $.7$ & 16 & 8 \\
 &  &  &  &  & F & $0.2$--$2.0$ &  & $25$ & $.7$ & $<$ & $12$ & $.6$ &  & $-4$ & $.6$ & $<$ & $-17$ & $.7$ & $<$ & $-13$ & $.2$ & 16 & 8 \\
 & \object{14542923$+$1606039 Bab} & M8.5 & $-3$ & $.17$ & Q & $0.5$--$8.0$ &  & $25$ & $.9$ & $<$ & $13$ & $.7$ &  & $-4$ & $.5$ & $<$ & $-16$ & $.8$ & $<$ & $-12$ & $.2$ & 19 & 8 \\
 &  &  &  &  & F & $0.5$--$8.0$ &  & $27$ & $.6$ & $<$ & $13$ & $.7$ &  & $-2$ & $.8$ & $<$ & $-16$ & $.8$ & $<$ & $-13$ & $.9$ & 19 & 8 \\
\P & \object{18353790$+$3259545} & M8.5 & $-3$ & $.52$ & -- & $0.2$--$2.0$ & $<$ & $24$ & $.5$ &  & $13$ & $.3$ & $<$ & $-5$ & $.6$ &  & $-16$ & $.8$ & $>$ & $-11$ & $.3$ & 16 & 16 \\
 & \object{01095117$-$0343264} & M9 & $-3$ & $.48$ & Q & $0.2$--$2.0$ &  & $25$ & $.2$ & $<$ & $12$ & $.5$ &  & $-4$ & $.9$ & $<$ & $-17$ & $.6$ & $<$ & $-12$ & $.7$ & $\star$ & $\star$ \\
 &  &  &  &  & F & $0.2$--$2.0$ &  & $26$ & $.1$ & $<$ & $12$ & $.5$ &  & $-4$ & $.0$ & $<$ & $-17$ & $.6$ & $<$ & $-13$ & $.6$ & $\star$ & $\star$ \\
 & \object{03393521$-$3525440} & M9 & $-3$ & $.81$ & Q & $0.1$--$10.0$ & $<$ & $24$ & $.0$ &  & $12$ & $.3$ & $<$ & $-5$ & $.7$ &  & $-17$ & $.4$ & $>$ & $-11$ & $.7$ & 20 & 8 \\
 &  &  &  &  & F & $0.1$--$10.0$ &  & $25$ & $.6$ &  & $13$ & $.9$ &  & $-4$ & $.2$ &  & $-15$ & $.9$ &  & $-11$ & $.7$ & 20 & 8 \\
 & \object{08533619$-$0329321} & M9 & $-3$ & $.52$ & Q & $0.3$--$0.8$ &  & $26$ & $.5$ & $<$ & $12$ & $.8$ &  & $-3$ & $.5$ & $<$ & $-17$ & $.2$ & $<$ & $-13$ & $.7$ & 21 & 22 \\
 &  &  &  &  & F & $0.1$--$2.4$ &  & $27$ & $.6$ & $<$ & $12$ & $.8$ &  & $-2$ & $.5$ & $<$ & $-17$ & $.2$ & $<$ & $-14$ & $.7$ & 23 & 22 \\
 & \object{10481463$-$3956062} & M9 & $-3$ & $.54$ & -- & $0.2$--$2.0$ &  & $25$ & $.1$ &  & $12$ & $.4$ &  & $-4$ & $.9$ &  & $-17$ & $.6$ &  & $-12$ & $.7$ & 26 & 8 \\
 & \object{14284323$+$3310391} & M9 & $-3$ & $.63$ & -- & $0.1$--$2.4$ & $<$ & $25$ & $.5$ & $<$ & $13$ & $.1$ & $<$ & $-4$ & $.4$ & $<$ & $-16$ & $.9$ &  & &  & 27 & 8 \\
\P & \object{15010818$+$2250020} & M9 & $-3$ & $.67$ & Q & $0.3$--$2.0$ &  & $24$ & $.9$ &  & $13$ & $.4$ &  & $-5$ & $.0$ &  & $-16$ & $.5$ &  & $-11$ & $.5$ & 28 & 28 \\
\P &  &  &  &  & F & $0.3$--$2.0$ &  & $27$ & $.3$ &  & $14$ & $.5$ &  & $-2$ & $.6$ &  & $-15$ & $.4$ &  & $-12$ & $.9$ & 28 & 28 \\
\P & \object{00242463$-$0158201} & M9.5 & $-3$ & $.49$ & -- & $0.2$--$2.0$ & $<$ & $25$ & $.1$ & $<$ & $12$ & $.9$ & $<$ & $-4$ & $.9$ & $<$ & $-17$ & $.2$ &  & &  & 30 & 30 \\
 & \object{00274197$+$0503417} & M9.5 & $-3$ & $.67$ & -- & $0.3$--$8.0$ & $<$ & $26$ & $.2$ & $<$ & $14$ & $.7$ & $<$ & $-3$ & $.7$ & $<$ & $-15$ & $.2$ &  & &  & 26 & 22 \\
\P & \object{07464256$+$2000321 AB} & L0 & $-3$ & $.36$ & Q & $0.2$--$2.0$ & $<$ & $25$ & $.2$ &  & $13$ & $.6$ & $<$ & $-5$ & $.0$ &  & $-16$ & $.6$ & $>$ & $-11$ & $.6$ & 31 & 31 \\
 &  &  &  &  & F & $0.2$--$2.0$ & $<$ & $25$ & $.2$ &  & $15$ & $.4$ & $<$ & $-5$ & $.0$ &  & $-14$ & $.8$ & $>$ & $-9$ & $.8$ & 31 & 8 \\
\P & \object{06023045$+$3910592} & L1 & $-3$ & $.67$ & -- & $0.2$--$2.0$ & $<$ & $25$ & $.0$ & $<$ & $12$ & $.8$ & $<$ & $-4$ & $.9$ & $<$ & $-17$ & $.2$ &  & &  & 30 & 30 \\
\P & \object{13054019$-$2541059 AB} & L2 & $-3$ & $.56$ & -- & $0.1$--$10.0$ &  & $25$ & $.4$ & $<$ & $12$ & $.8$ &  & $-4$ & $.7$ & $<$ & $-17$ & $.3$ & $<$ & $-12$ & $.6$ & 32 & 32 \\
\P & \object{05233822$-$1403022} & L2.5 & $-3$ & $.82$ & -- & $0.2$--$2.0$ & $<$ & $25$ & $.2$ & $<$ & $13$ & $.0$ & $<$ & $-4$ & $.5$ & $<$ & $-16$ & $.8$ &  & &  & 30 & 30 \\
\P & \object{00361617$+$1821104} & L3.5 & $-3$ & $.99$ & Q & $0.2$--$8.0$ & $<$ & $24$ & $.9$ &  & $13$ & $.4$ & $<$ & $-4$ & $.7$ &  & $-16$ & $.2$ & $>$ & $-11$ & $.6$ & 33 & 33 \\
 &  &  &  &  & F & $0.2$--$8.0$ & $<$ & $24$ & $.9$ &  & $13$ & $.8$ & $<$ & $-4$ & $.7$ &  & $-15$ & $.8$ & $>$ & $-11$ & $.1$ & 33 & 22 \\
 & \object{12281523$-$1547342 AB} & L5 & $-3$ & $.93$ & -- & $0.3$--$8.0$ & $<$ & $26$ & $.6$ & $<$ & $13$ & $.6$ & $<$ & $-3$ & $.0$ & $<$ & $-16$ & $.0$ &  & &  & 26 & 22 \\
 & \object{15074769$-$1627386} & L5 & $-4$ & $.23$ & -- & $0.2$--$8.0$ & $<$ & $24$ & $.8$ & $<$ & $12$ & $.6$ & $<$ & $-4$ & $.6$ & $<$ & $-16$ & $.8$ &  & &  & 33 & 22
\enddata
\tablecomments{Rows marked with a pilcrow (\P) indicate simultaneous radio and X-ray
observations. Col. (4) is the state of the source: quiescent (Q), flaring (F),
or indeterminate/unknown (--). \lx\ has been normalized to the 0.2--2~keV
bandpass as described in the text.}
\tablerefs{Columns are (X), X-ray flux; (R), radio flux. [$\star$]: this work, [1] \citet{hks94}, [2] \citet{thegctp}, [3] \citet{khm91}, [4] \citet{ltsr09}, [5] \citet[in preparation]{dracut}, [6] \citet{crl+03}, [7] \citet{sfg95}, [8] \citet{mbr12}, [9] \citet{thethirdgj}, [10] \citet{gsbf93}, [11] \citet{cpbd+05}, [12] \citet{gc09}, [13] \citet{crk+07}, [14] \citet{ssml06}, [15] \citet{fcs03}, [16] \citet{bbg+08}, [17] \citet{zolp+04}, [18] \citet{s04}, [19] \citet{smf+06}, [20] \citet{rbmb00}, [21] \citet{rs08}, [22] \citet{b02}, [23] \citet{sl02}, [24] \citet{rb10}, [25] \citet{rck+08}, [26] \citet{sab+12}, [27] \citet{fgsb93}, [28] \citet{bgg+08}, [29] \citet{rhg95}, [30] \citet{bbf+10}, [31] \citet{brpb+09}, [32] \citet{aob+07}, [33] \citet{brr+05}}
\end{deluxetable}

\end{turnpage}
\clearpage
\pwifjournal\else
  \global\pdfpageattr\expandafter{\the\pdfpageattr/Rotate 90}
\fi

\end{document}